\documentclass{aastex}

\newcommand\kms    {\ifmmode{{\rm ~km~s}^{-1}~}\else{~km~s$^{-1}~$}\fi}

\newcommand\lo     {~$L_{\odot}$}
\newcommand\Lsun   {~$L_{\odot}$}
\newcommand\mo     {~$M_{\odot}$}
\newcommand\Msun   {~$M_{\odot}$}

\newcommand\couc    {CO(1$\rightarrow$0)}
\newcommand\trecouc {$^{13}$CO(1$\rightarrow$0)}
\newcommand\csdu    {CS(2$\rightarrow$1)}
\newcommand\cstd    {CS(3$\rightarrow$2)}
\newcommand\ctcstd  {C$^{34}$S(3$\rightarrow$2)}
\newcommand\siodu   {SiO(2$\rightarrow$1)}
\newcommand\siotd   {SiO(3$\rightarrow$2)}
\newcommand{\jj}[2]{\mbox{$(#1 \rightarrow #2)$}}
\newcommand{\Jjj}[2]{\mbox{$J = #1 \rightarrow #2$}}
\newcommand{\jk}[2]{\mbox{$(#1_{k} \rightarrow #2_{k})$}}

\newcommand{\jkk}[3]{\mbox{$(#1_{#3} \rightarrow #2_{#3})$}}
\newcommand{\jkks}[3]{\mbox{$#1_{#3} \rightarrow #2_{#3}$}}
\newcommand{\Jjkk}[3]{\mbox{$J_{k} = #1_{#3} \rightarrow #2_{#3}$}}

\begin{document}
\title{Four highly luminous massive star forming regions in the 
Norma Spiral Arm.} 


\title{I. Molecular gas and dust observations} 

\author{Guido Garay\altaffilmark{1}, Diego Mardones\altaffilmark{1,2},
Leonardo Bronfman\altaffilmark{1}, Jorge May\altaffilmark{1}, 
Luis Chavarr\'\i a\altaffilmark{1}, and Lars-{\AA}ke Nyman\altaffilmark{3}}


\altaffiltext{1}{Departamento de Astronom\'{\i}a, Universidad de Chile,
    Camino del Observatorio 1515, Las Condes, Santiago, Chile}
\altaffiltext{2}{Centro de Radioastronom\'\i a y Astrof\'\i sica, UNAM,
Apdo. Postal 3-72, Morelia, Michoac\'an, 58089 M\'exico}
\altaffiltext{3}{European Southern Observatory, Casilla 19001, Santiago, Chile}

\begin{abstract}

We report molecular line and dust continuum observations, made with the SEST 
telescope, towards four young high-mass star forming regions 
associated with highly luminous (${\cal L}> 6\times10^5{\cal L}_{\odot}$) IRAS 
sources (15290-5546, 15502-5302, 15567-5236 and 16060-5146). Molecular emission 
was mapped in three lines of CS (J=2$\rightarrow$1, 3$\rightarrow$2 and 
5$\rightarrow$4), two lines of SiO (J=2$\rightarrow$1 and 3$\rightarrow$2), 
two rotational transitions of CH$_3$OH (J$_k$=3$_k\rightarrow2_k$ 
and $2_k\rightarrow1_k$), and in the C$^{34}$S(J=3$\rightarrow$2) line. In 
addition, single spectra at the peak position were taken in the 
CO(J=1$\rightarrow$0), $^{13}$CO(J=1$\rightarrow$0) and 
C$^{18}$O(J=1$\rightarrow$0) lines.
We find that the luminous star forming regions are associated 
with molecular gas and dust structures with radii of typically 0.5 pc, 
masses of $\sim 5\times10^3$ M$_\odot$, column densities of $\sim 5\times10^{23}$ 
cm$^{-2}$, molecular hydrogen densities of typically $\sim2\times10^{5}$ 
cm$^{-3}$ and dust temperatures of $\sim 40$ K. The 1.2 mm dust 
continuum observations further indicate that the cores are centrally condensed, 
having radial density profiles with power-law indices in the range $1.6-1.9$.   
We find that under these conditions dynamical friction by the gas plays an 
important role in the migration of high-mass stars towards the
central core region,  providing an explanation for the observed stellar mass 
segregation within the cores. 

The CS profiles show two distinct emission components: a bright component, with 
line widths of typically 5 \kms (FWHM), and a weaker and wider velocity 
component, which extends up to typically $\pm$13 \kms from the ambient cloud 
velocity. The SiO profiles also show emission from both components, but the 
intensity of the pedestal feature relative to that of the bright component is 
stronger than for CS. The narrow SiO component is likely to trace warm ambient gas 
close to the recently formed massive stars, whereas the high velocity emission 
indicates mass outflows produced by either the expansion of the HII regions, 
stellar winds, and/or collimated outflows. We find that the abundances of CS, 
CH$_3$OH and SiO, relative to H$_2$, in the warm ambient gas of the massive cores 
are typically $4\times10^{-8}$, $6\times10^{-9}$, and $5\times10^{-11}$, 
respectively. 

\end{abstract}  


\vfill\eject

\section{INTRODUCTION}

During the last decades several observational efforts have been carried out to
determine the characteristics of the molecular gas and dust associated with
massive star forming regions. Surveys of molecular emission in high density
tracers (Plume et al. 1992; Juvela 1996; Plume et al. 1997; Shirley et al. 2003)
and of dust continuum emission (Beuther et al. 2002; Mueller et al. 2002;
Fa\'undez et al. 2004; Williams et al. 2004; Garay et al. 2007), made with 
single dish telescopes, show that the maternities of high-mass stars are regions 
of molecular gas with distinctive physical parameters, with typical radii of 0.4 
pc, densities of $5\times10^5$ cm$^{-3}$ and masses of $4\times10^3$\Msun. The 
dust observations also show that these regions have dust temperatures of typically
$32$ K, indicating that a luminous energy source has been already formed
inside them.  These structures, also refereed as massive and dense cores,
have typically column densities of $3\times10^{23}$ cm$^{-2}$ (Garay et al. 2007) 
which make them very dark at optical wavelengths and even dark at
infrared wavelengths (Menten et al. 2005, Rathborne et al. 2006).
The IRAS luminosity of most regions in these studies are in the 
range $1\times10^4 - 4\times10^5$ \Lsun.  

In order to investigate possible differences, in the physical and chemical 
characteristics as well as in the stellar population, between massive 
and dense cores of different luminosities, we have carried out a multi-wavelength 
study of four highly luminous IRAS sources in 
the southern hemisphere selected from the list of IRAS point sources 
with colors of UC H{\small II} regions and detected in 
CS(2$\rightarrow$1) emission compiled by Bronfman et al. (1996). The
observed objects are listed in Table~\ref{tbl-obssources}. Cols. 2, 3, and 4
give, respectively, the name, right ascension, and declination of the associated
IRAS source.  The kinematic distance to the objects is given in col. 5.
The four objects are located in the tangent region of the Norma spiral
arm and therefore there is little ambiguity in the kinematic determination of their
distances; their $v_{\small LSR}$'s are nearly the maxima allowed by pure circular
motion in their lines of sight. The total far infrared luminosities, $L_{FIR}$,
are given in col.6. The objects have $L_{FIR}$ greater than 
$6\times10^5$ $L_{\odot}$  and their CS(2$\rightarrow$1) line profiles show broad
molecular line wing emission suggesting the presence of powerful molecular outflows.

Here we report molecular line and dust continuum observations towards these four 
regions. The aim of these observations is to undertake a detailed study of the 
molecular gas and dust content of luminous massive and dense cores in order to 
determine their physical conditions, kinematics, and chemistry. Molecular emission 
was mapped in a number of species that are commonly used as diagnostics of the 
conditions of massive and dense cores: carbon monosulfide, silicon monoxide, 
and methanol (e.g., Zinchenko et al. 1995; Juvela 1996). Thermal emission 
of SiO is both a good probe of high velocity gas in regions of massive star 
formation (Downes et al. 1982) and a tracer of high-temperature chemistry 
(Ziurys et al. 1989). The methanol abundance is highly enhanced in the hot gas 
surrounding infrared sources (Menten et al. 1988) as well as in shocked gas 
(Liechti \& Walmsley 1997), thus thermal methanol emission is an excellent 
probe of the environments of high-mass YSO's. The CS and dust continuum 
observations are intended to determine the density structure of the cores. In a 
subsequent paper (Chavarr\'\i a et al. 2009; Paper II) we will 
present the characteristics of the stellar content within these regions. 

\section{OBSERVATIONS}

The molecular line and dust continuum observations were carried out using 
the 15-m Swedish-ESO Submillimetre Telescope (SEST) located on La Silla, Chile. 

\subsection{Molecular lines}

The molecules, transitions, and frequencies observed and the instrumental 
parameters are summarized in Table~\ref{tbl-obspar}. 

The CS($2\rightarrow1$) observations were made between December 1990 and January 
1992. During this period the telescope was equipped with a cooled 
Schottky mixer and an acousto-optical spectrometer with 2000$\times$43 KHz 
channels. The typical system temperature above the atmosphere was 500 K. The 
CS($2\rightarrow1$) line emission was mapped with spacings of 45\arcsec\ within 
regions of at least 135\arcsec$\times$135\arcsec. The observations were done in 
the frequency switch mode with a frequency throw of 15 MHz.

The SiO($2\rightarrow1$), SiO($3\rightarrow2$), CH$_3$OH($2_k\rightarrow1_k$), 
CH$_3$OH($3_k\rightarrow2_k$), CS($3\rightarrow2$) and C$^{34}$S($3\rightarrow2$) 
observations were made during March of 1997. In this epoch the telescope was 
equipped with SIS receivers, and it was possible to simultaneously observe lines 
at 2 and 3 mm wavelengths. Single-sideband receiver temperatures were typically 
120 K for both receivers. As backend we used high resolution acousto-optical 
spectrometers providing a channel separation of 43 kHz and a total bandwidth of 
43 MHz. The emission in the above lines was mapped within regions of at least 
90\arcsec $\times$ 90\arcsec, with 30\arcsec\ spacings. The observations were 
performed in dual beam-switching mode, with a beam separation of 
11\arcmin47\arcsec\ in azimuth. The integration times on source per position and 
the resulting rms noise in antenna temperature are given in Table~\ref{tbl-tinrms}. 
Within the available bandwidths, three rotational lines of CH$_3$OH could be 
observed at 2 mm (\jkks{3}{2}{0}\ A$^+$, \jkks{3}{2}{-1}\ E, and 
\jkks{3}{2}{0}\ E lines) and four rotational lines at 3 mm (\jkks{2}{1}{1}\ E, 
\jkks{2}{1}{0}\ E, \jkks{2}{1}{0}\ A$^+$, and \jkks{2}{1}{-1}\ E lines). 
During March 2001 we further observed the SiO\jj{5}{4} (217104.935 GHz) and 
CH$_3$OH($5_{-1}\rightarrow4_0$) (84521.21 GHz) lines 
toward the center of the G330.949-0.174 core. The receivers and backends
were the same as described above. Single-sideband receiver temperatures 
were 1200 K at 217 GHz and 190 K at 85 GHz.

The CS($5\rightarrow4$) observations of G324.201, G329.337, and G330.949 
were made between September 1991 and January 1992. In this epoch the 
telescope was equipped with a cooled Schottky mixer, and an acousto-optical 
spectrometer with 2000$\times$43 KHz channels; however, a setup of 
1000$\times$86 KHz resolution elements was used.
The typical system temperature above the atmosphere was 1850 K. The 
CS($5\rightarrow4$) line emission was mapped with spacings of 
15\arcsec\ within regions of at least 60\arcsec$\times$45\arcsec. The 
observations were done in the frequency switch mode with a
frequency throw of 15 MHz.  The CS($5\rightarrow4$) observations of G328.307 
were made in May 2001, with an SIS receiver and the acousto-optical spectrometer 
with 2000$\times$43 KHz channels.  Typical system temperature above the 
atmosphere was of 450 K. Observations were done in dual beam switching mode, 
with a beam separation of 11\arcmin47\arcsec\ in azimuth,
in a grid of 9 positions with spacings of 45\arcsec.

Observations of the CO, $^{13}$CO, and C$^{18}$O emission in the J=1$\rightarrow$0
lines were made during May, 2000 and June, 2001. Spectra were obtained only toward
the peak position of the CS emission in each region. These observations were
performed in the position switched mode, with OFF positions taken from Bronfman
et al. (1989). The goal of these observations was to obtain the data needed to
determine CO column densities, which in turn will permit to derive abundance
enhancements relative to CO.

\subsection{Dust continuum}

The 1.2-mm continuum observations were made
using the 37-channel SEST Imaging Bolometer Array (SIMBA) during October, 2001
and August, 2003. The passband of the bolometers has an equivalent width
of $90$ GHz and is centered at $250$ GHz. The HPBW of a single element is
24\arcsec\ and the separation between elements on the sky is 44\arcsec. We
observed in the fast mapping mode, using a scan speed of 80\arcsec\ s$^{-1}$.
Each observing block consisted of 50 scan lines in azimuth of length 800\arcsec\
and separated in elevation by 8\arcsec, giving a map size of 400\arcsec\
in elevation. This block required $\sim$ 15 minutes of observing time. We
observed $2-4$ blocks per source, achieving rms noise levels of typically $50$
mJy beam$^{-1}$. The data were reduced according to a standard procedure
using the software package MOPSI, which included baseline subtraction and
rejection of correlated sky-noise. Flux calibration was performed using a
sky-opacity correction and a counts-to-flux conversion factor derived from maps
of Uranus. Uncertainties in the absolute calibration and pointing accuracy are
estimated at $20\%$ and 3\arcsec, respectively.

\section{RESULTS}

\subsection{Molecular emission}

\begin{figure}
\epsscale{0.82}
\plotone{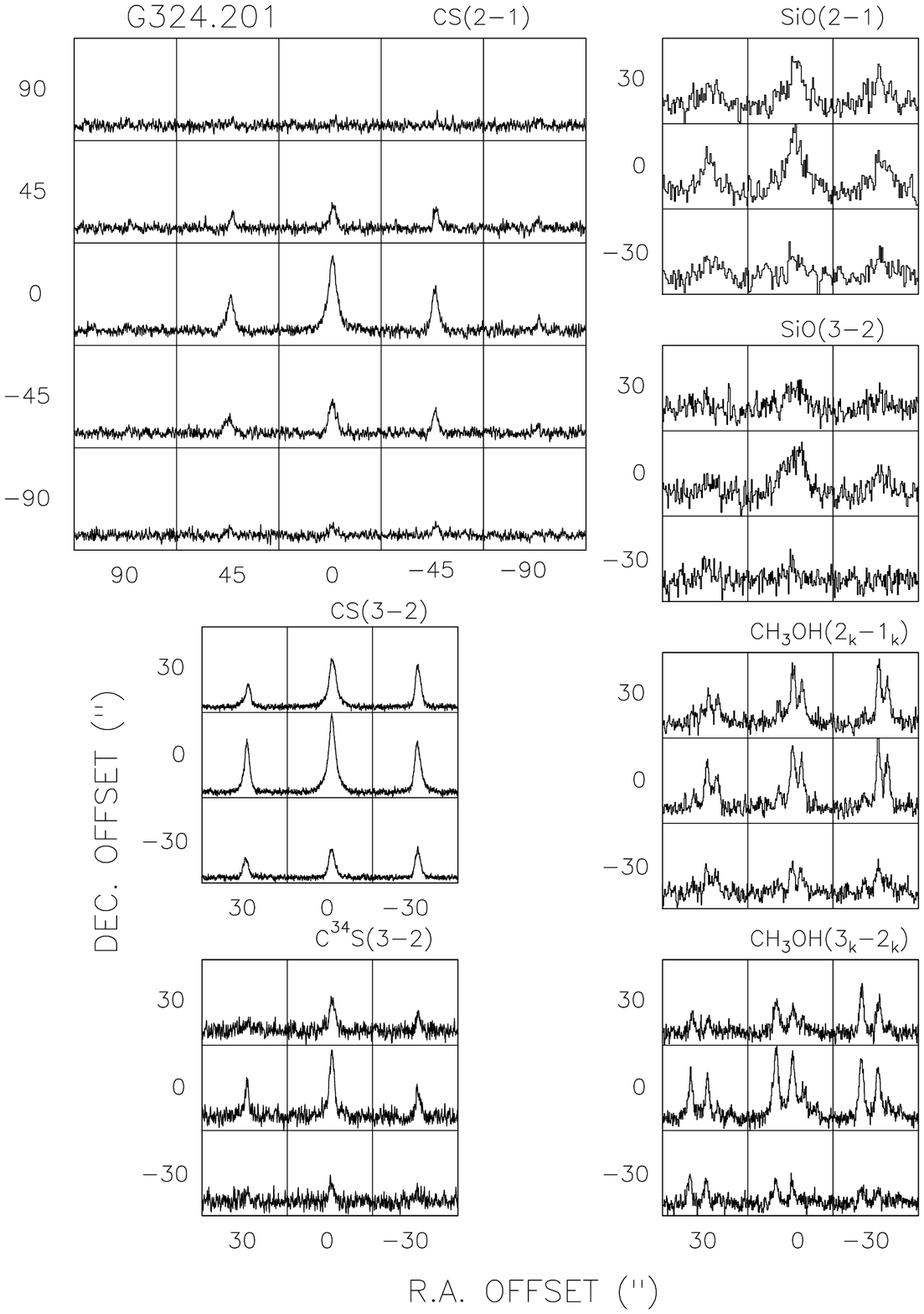}
\caption
{\baselineskip2.0pt
Spectral maps of the molecular line emission from G324.201+0.119.
The grid spacing is 30\arcsec\ for all lines, except \csdu\ which is 45\arcsec.  
Offsets are from the reference position at $\alpha_{2000} =
15^{\rm h}32^{\rm m}53\fs2, \delta_{2000} = -55\arcdeg56\arcmin13\arcsec$. 
The velocity scale ranges from $-120$ to $-60$ \kms.
The antenna temperature scale ranges from $-0.5$ to 3.5 K for \csdu, 
$-0.25$ to 3.6 K for \cstd, $-0.2$ to 1.0 K for \ctcstd, $-0.03$ to 0.12 K for 
\siodu, $-0.06$ to 0.18 K for \siotd, $-0.1$ to 0.45 K for CH$_3$OH($2_k-1_k$), 
and $-0.1$ to 0.60 K for CH$_3$OH($3_k-2_k$).
\label{fig-specmapg324}}
\end{figure}

\begin{figure}
\epsscale{0.85}
\plotone{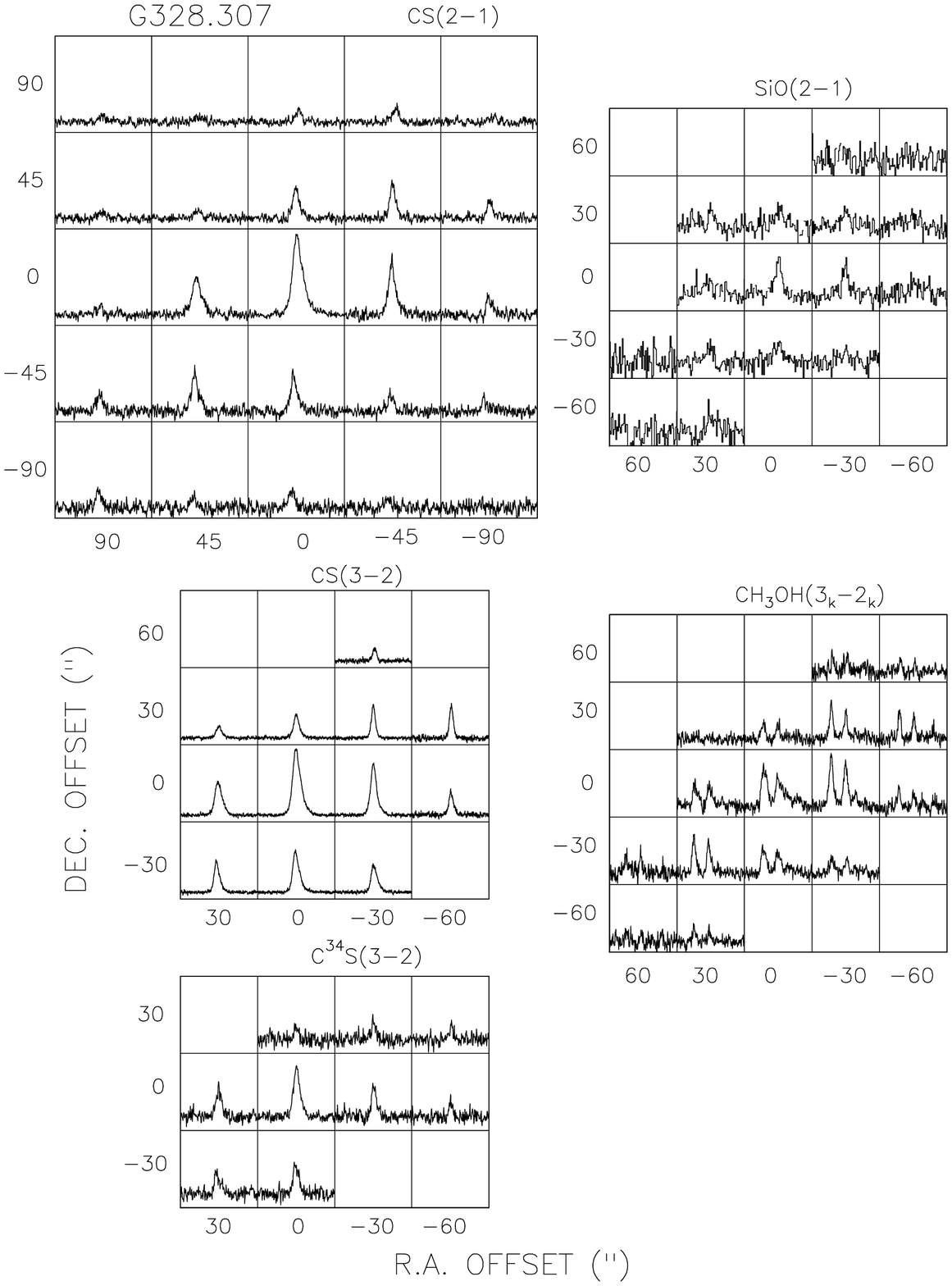}
\caption
{\baselineskip3.0pt
Spectral maps of the molecular line emission from G328.307+0.423. 
The grid spacing is 30\arcsec\ for all lines, except \csdu\ which is 45\arcsec.  
Offsets are from the reference position at $\alpha_{2000} =
15^{\rm h}54^{\rm m}06\fs3, \delta_{2000} = -53\arcdeg11\arcmin38\arcsec$. 
The velocity scale ranges from $-120$ to $-65$ \kms.
The antenna temperature scale ranges from $-0.5$ to 4.0 K for \csdu, 
$-0.5$ to 5.0 K for \cstd, $-0.2$ to 0.9 K for \ctcstd, $-0.04$ to 0.13 K for 
\siodu, and $-0.1$ to 0.55 K for CH$_3$OH($3_k-2_k$).
\label{fig-specmapg328}}
\end{figure}

\begin{figure}
\epsscale{0.85}
\plotone{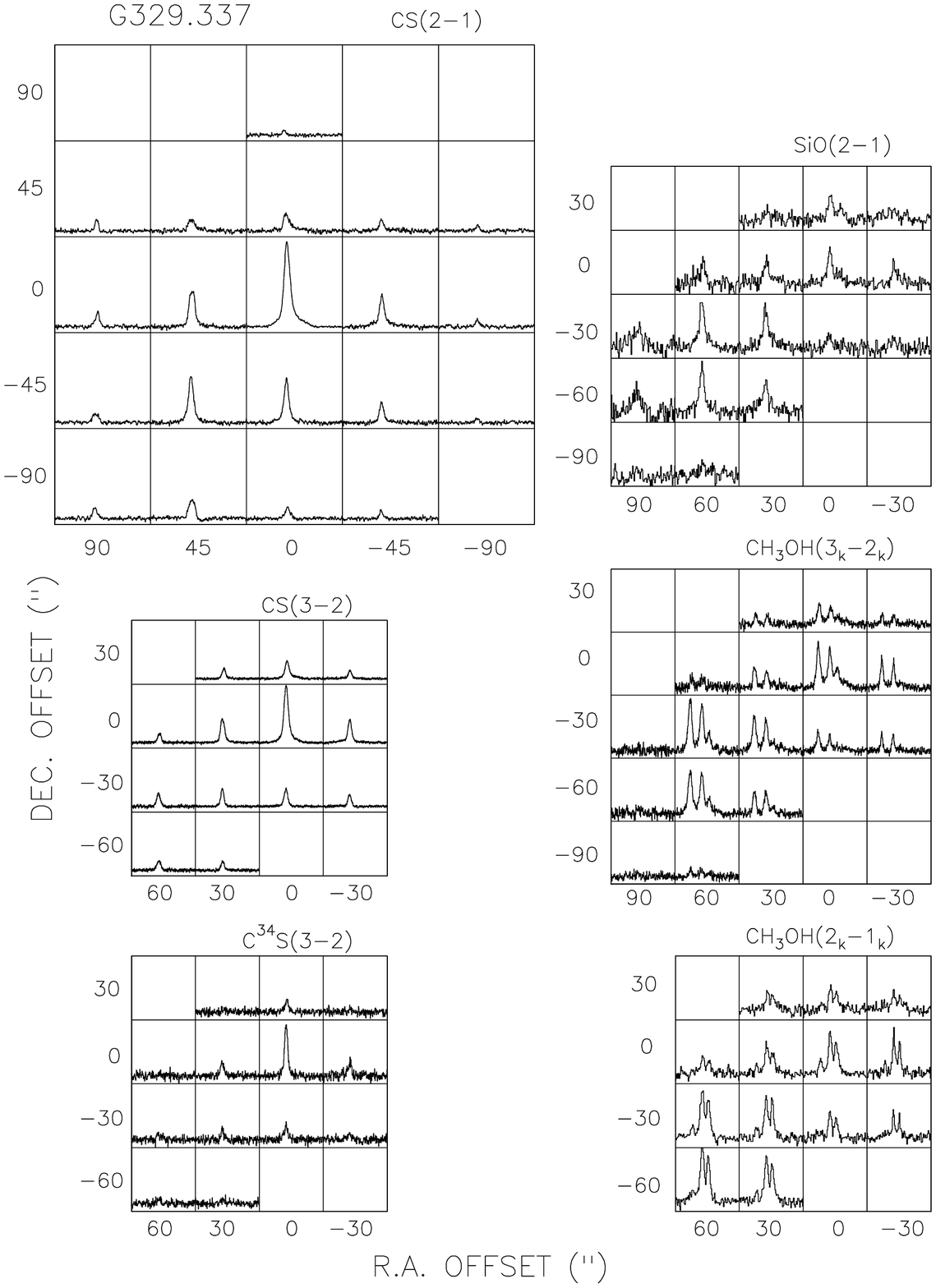}
\caption
{\baselineskip3.0pt
Spectral maps of the molecular line emission from G329.337+0.147. 
The grid spacing is 30\arcsec\ for all lines, except \csdu\ which is 45\arcsec.  
Offsets are from the reference position at $\alpha_{2000} =
16^{\rm h}00^{\rm m}33\fs4, \delta_{2000} = -52\arcdeg44\arcmin45\arcsec$. 
The velocity scale ranges from $-135$ to $-70$ \kms.
The antenna temperature scale ranges from $-0.3$ to 4.5 K for \csdu, 
$-0.6$ to 6.1 K for \cstd, $-0.2$ to 1.4 K for \ctcstd, $-0.04$ to 0.21 K for 
\siodu, $-0.12$ to 0.90 K for CH$_3$OH($3_k-2_k$), and
$-0.12$ to 0.65 for CH$_3$OH($2_k-1_k$).
\label{fig-specmapg329}}
\end{figure}

\begin{figure}
\epsscale{0.78}
\plotone{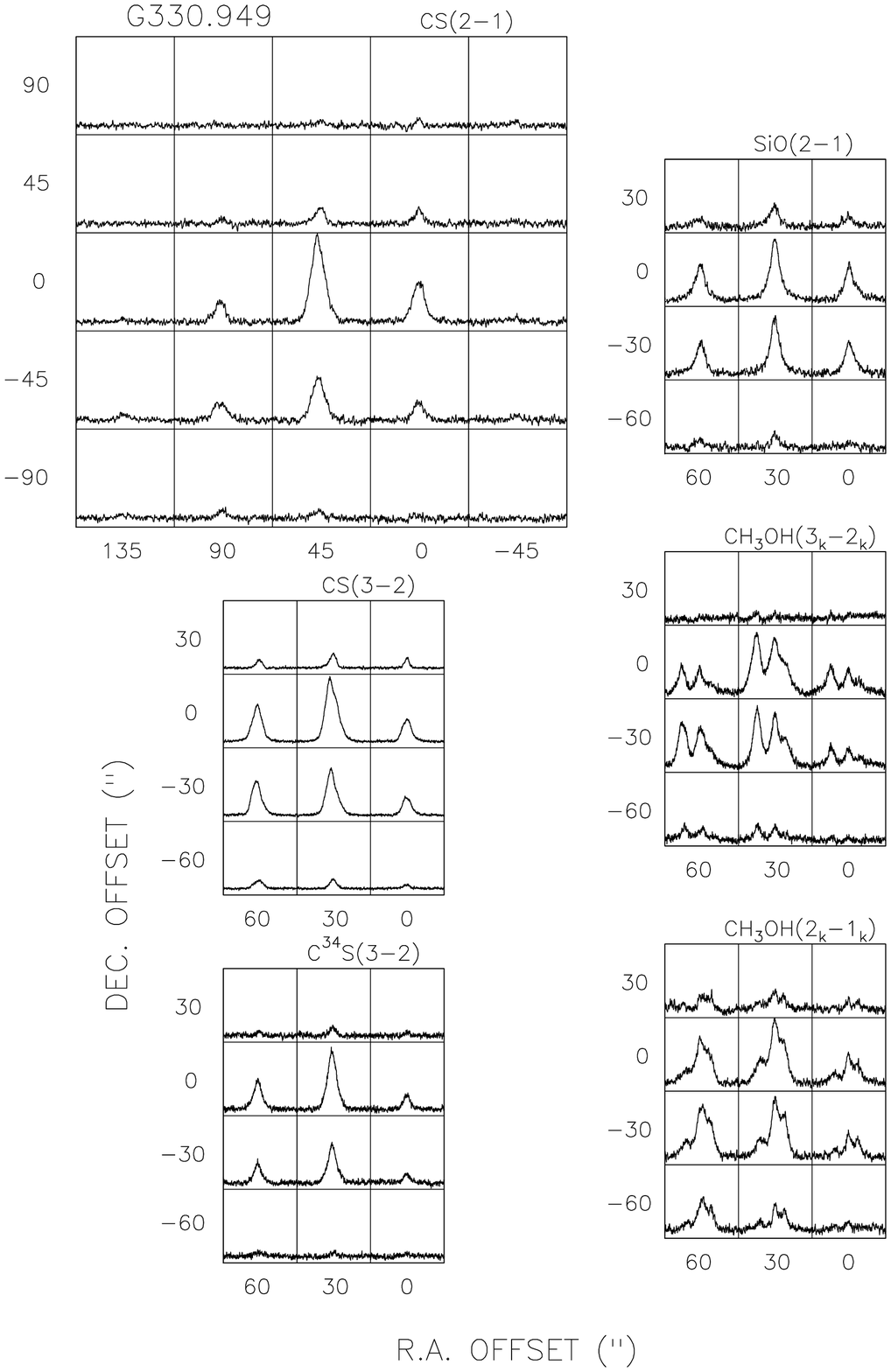}
\caption
{\baselineskip2.0pt
Spectral maps of the molecular line emission from G330.949-0.174.
The grid spacing is 30\arcsec\ for all lines, except \csdu\ which is 45\arcsec.  
Offsets are from the reference position at $\alpha_{2000} =
16^{\rm h}09^{\rm m}49\fs1, \delta_{2000} = -51\arcdeg54\arcmin45\arcsec$. 
The velocity scale ranges from $-115$ to $-65$ \kms.
The antenna temperature scale ranges from $-0.5$ to 4.8 K for \csdu, 
$-0.6$ to 6.1 K for \cstd, $-0.2$ to 2.0 K for \ctcstd, $-0.07$ to 0.75 K for 
\siodu, $-0.15$ to 1.5 K for CH$_3$OH($3_k-2_k$), and
$-0.15$ to 1.1 for CH$_3$OH($2_k-1_k$).
\label{fig-specmapg330}}
\end{figure}

Figs.~\ref{fig-specmapg324} to \ref{fig-specmapg330} show spectral maps of the 
CS(2$\rightarrow$1), CS(3$\rightarrow$2), C$^{34}$S(3$\rightarrow$2), 
SiO(2$\rightarrow$1), SiO(3$\rightarrow$2), CH$_3$OH(3$_k\rightarrow2_k$), and 
CH$_3$OH(2$_k\rightarrow1_k$) line emission observed toward G324.201+0.119, 
G328.307+0.423, G329.337+0.147 and G330.949-0.174, respectively. The spacing is 
30\arcsec\ for all transitions, except CS(2$\rightarrow$1) for which the spacing is 
45\arcsec. Offsets are from the reference position given in columns 3 and 4 of 
Table~\ref{tbl-obssources}. Table~\ref{tbl-linepar} gives the parameters 
of the line emission observed at the peak position (cols. 2 to 4) and of the 
spatially averaged emission (cols. 5 to 7), determined from Gaussian fits
to the respective line profiles.

Carbon monosulfide was detected in all observed transitions towards all sources.
The CS(2$\rightarrow$1) and CS(3$\rightarrow$2) line profiles indicate the 
presence of two emission components: (i) a bright, spatially extended, and 
relatively narrow velocity feature; and (ii) a weaker and broader velocity 
feature, which is usually spatially confined to the central region. 
The line widths of the narrow component emission, which most likely arises from 
the dense ambient gas, are typically about 5 \kms, a value considerably larger
than the thermal width expected for kinetic temperatures of $\sim$40 K, of 
$\sim0.2$ \kms,
implying that the line widths are dominated by non-thermal supersonic motions.
Whether the non-thermal motions are due to the effect of stellar winds or are
intrinsic to the initial conditions of the formation process of massive stars
is still an open question.

The velocity range of the broad wing emission (full width at zero power) 
in different transitions are given in Table~\ref{tbl-wingpar}. 
The broad emission, which has typically full widths at zero power of
$\sim26$ \kms, is mainly detected toward the peak position of the massive cores.
Although this emission most likely arises from outflowing gas,
the coarse angular resolution ($\sim30$\arcsec) of our observations does not allow
to determine its spatial distribution and degree of collimation. The nature of the
high velocity gas giving rise to the pedestal feature is thus difficult to assess.
Higher angular resolution observations are needed to investigate whether the
high velocity emission arises from a collimated bipolar outflow driven by the
recently formed massive stars or from an spherically symmetric structure possible
driven either by stellar winds or by the expansion of the HII region onto the 
ambient medium.

Silicon monoxide was detected toward the four observed high-mass star forming 
regions, the strongest emission being observed toward the G330.949-0.174
core. SiO lines are tracers of both the warm ambient gas surrounding young
luminous stars and of molecular outflows associated with newly formed massive
stars. Harju et al. (1998) have already shown that SiO emission is relatively easy
to detect toward regions of high-mass star formation, with the detection rate
increasing with the FIR luminosity.
The profiles of the SiO emission show a broad component of high velocity gas
superposed on a narrower component, whose central velocity is close to the velocity
of the ambient gas. The full widths at zero power (emission above a $3\sigma$
level) of the broad component range from 17.4 to 36.6 \kms, with a median of 26
\kms. The relative contribution to the observed intensity is different from source
to source. In G324.201 the SiO emission from the broad component dominates the
spectrum, whereas in G330.949 the narrower emission is stronger. We suggest that,
on the one hand, 
the narrow SiO emission arises from the warm ambient gas surrounding the recently
formed massive stars, and that this component results from the evaporation of
silicon compounds from grain mantles produced by the stellar radiation.
On the other hand, the broad SiO emission most likely indicates mass outflows from
a star or stars embedded within the massive core, and that this component results
from grain disruption by shock waves. Hence, the SiO emission traces both high
kinetic temperature ambient gas and outflowing shocked gas.

\begin{figure}
\epsscale{0.48}
\plotone{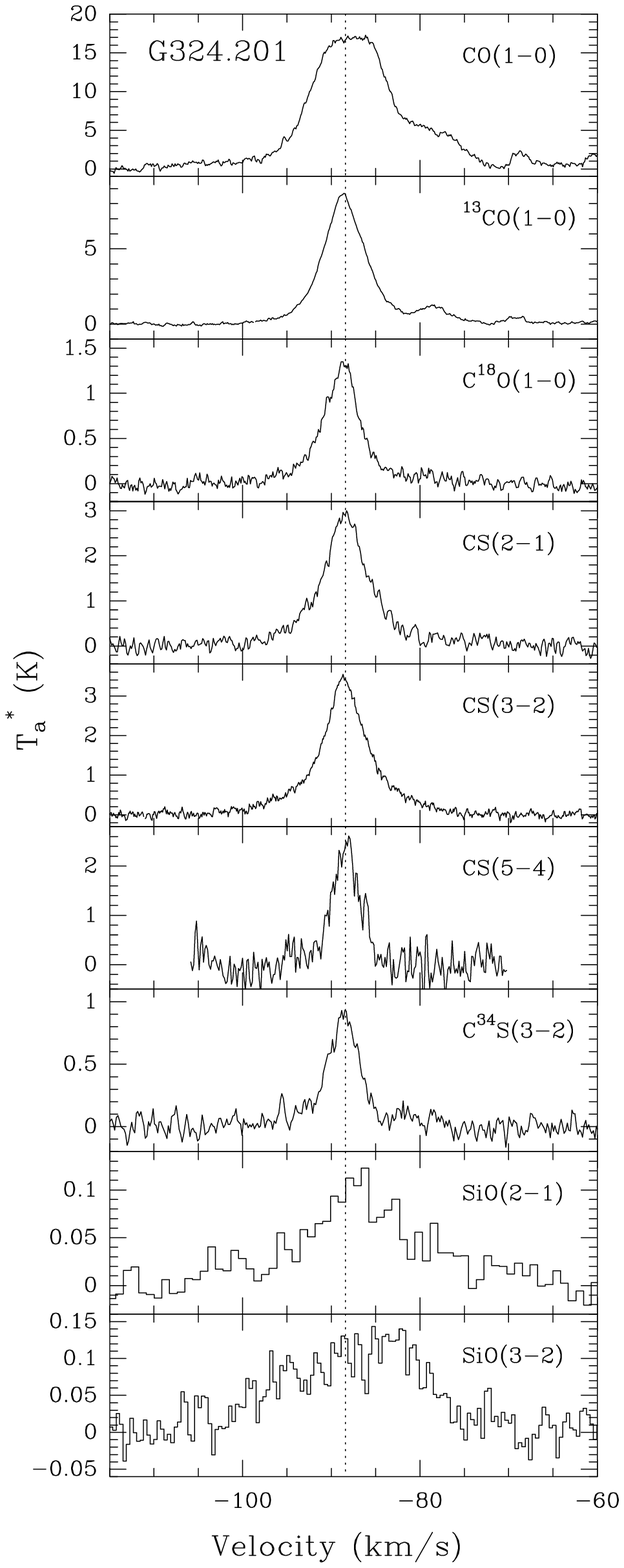}
\caption
{\baselineskip2.0pt
Spectra observed at the peak position of G324.201+0.119. 
Transitions are given in the upper right corner. The vertical
dotted line indicates the systemic velocity of the ambient gas. 
\label{fig-specpeakg324}}
\end{figure}

\begin{figure}
\epsscale{0.48}
\plotone{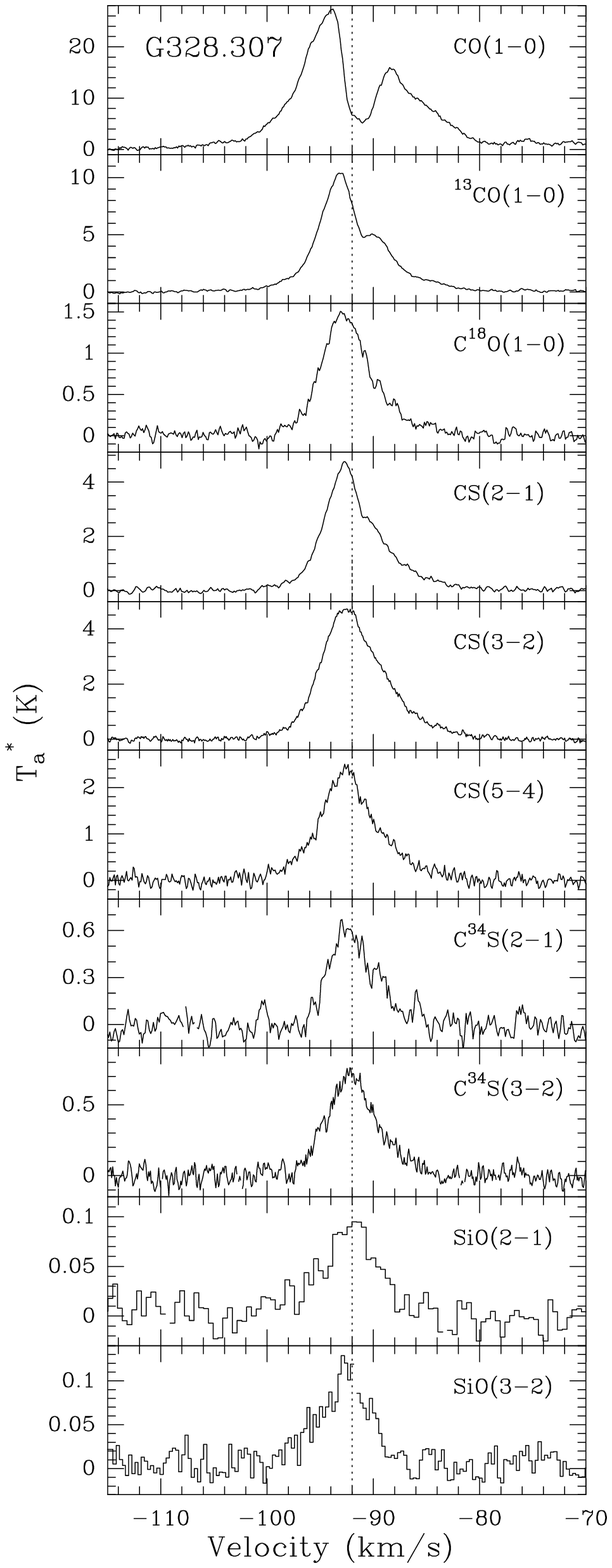}
\caption
{\baselineskip2.0pt
Spectra observed at the peak position of G328.307+0.423. 
Transitions are given in the upper right corner. The vertical
dotted line indicates the systemic velocity of the ambient gas. 
\label{fig-specpeakg328}}
\end{figure}

\begin{figure}
\epsscale{0.48}
\plotone{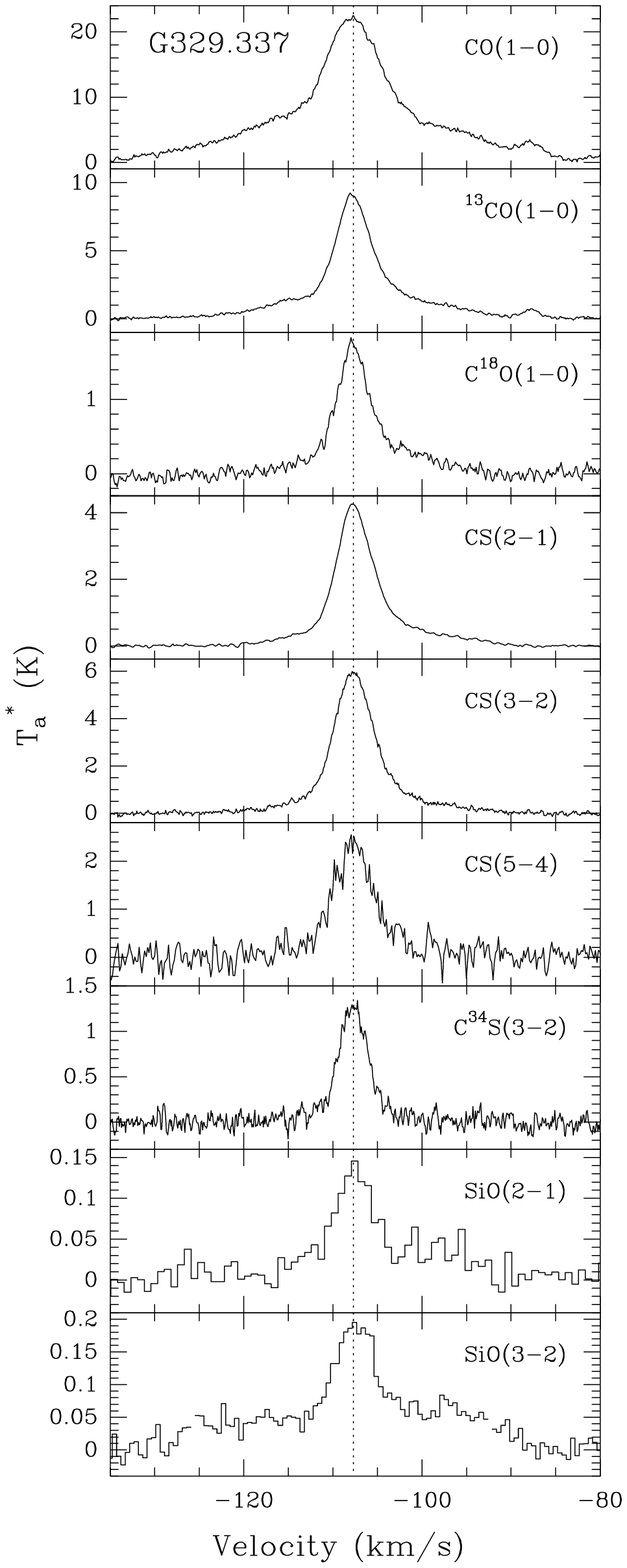}
\caption
{\baselineskip3.0pt
Spectra observed at the peak position of G329.337+0.147.
Transitions are given in the upper right corner. The vertical
dotted line indicates the systemic velocity of the ambient gas. 
\label{fig-specpeakg329}}
\end{figure}

\begin{figure}
\epsscale{0.48}
\plotone{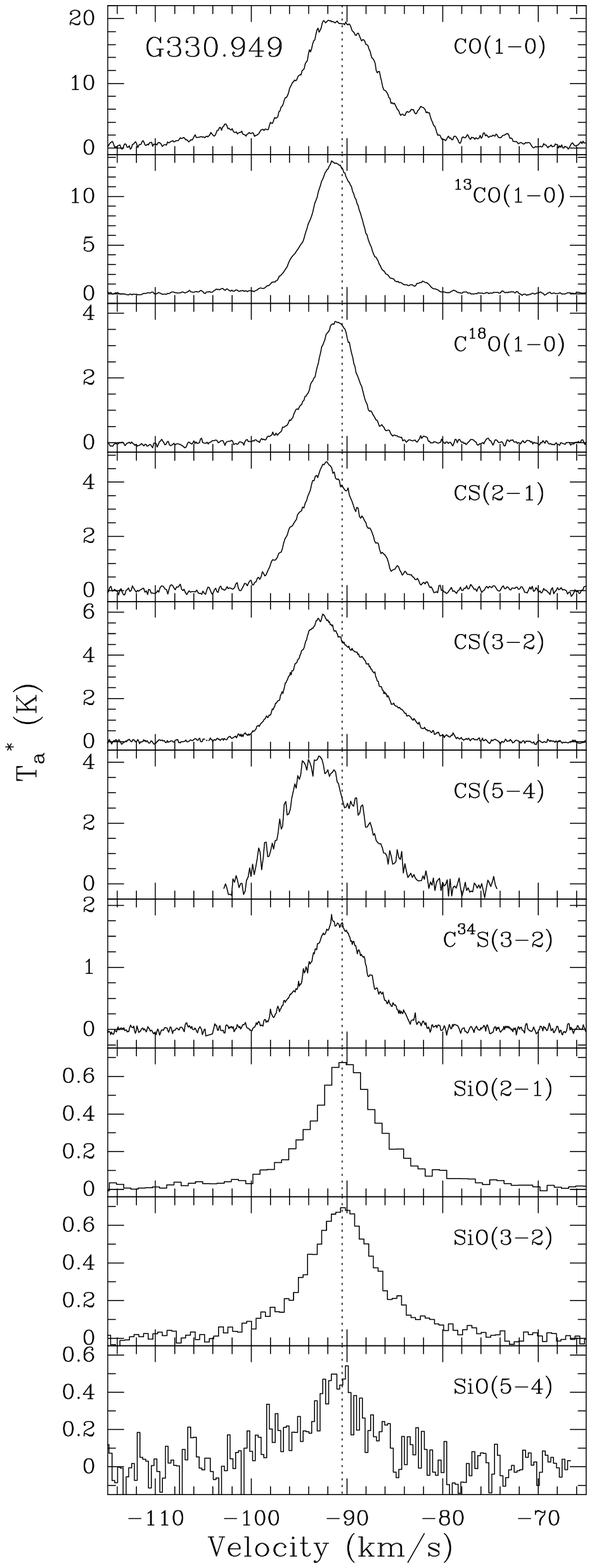}
\caption
{\baselineskip2.0pt
Spectra observed at the peak position of G330.949-0.174.
Transitions are given in the upper right corner. The vertical
dotted line indicates the systemic velocity of the ambient gas. 
\label{fig-specpeakg330}}
\end{figure}

The shape of the profiles are different from molecule to molecule depending on the 
relative intensities between the narrow and broad components. To illustrate in more
detail the differences between the line profiles, Figs.~\ref{fig-specpeakg324} 
to \ref{fig-specpeakg330} show the spectra of all molecular lines, except methanol, 
observed at the peak position of each region. The CS(5$\rightarrow$4) spectra 
corresponds to the average spectra over the 30\arcsec$\times$30\arcsec\ mapped 
region, and therefore it is directly comparable to the peak spectra observed in 
the other transitions with $\sim$30\arcsec\ beams. 

\begin{figure}
\epsscale{0.80}
\plotone{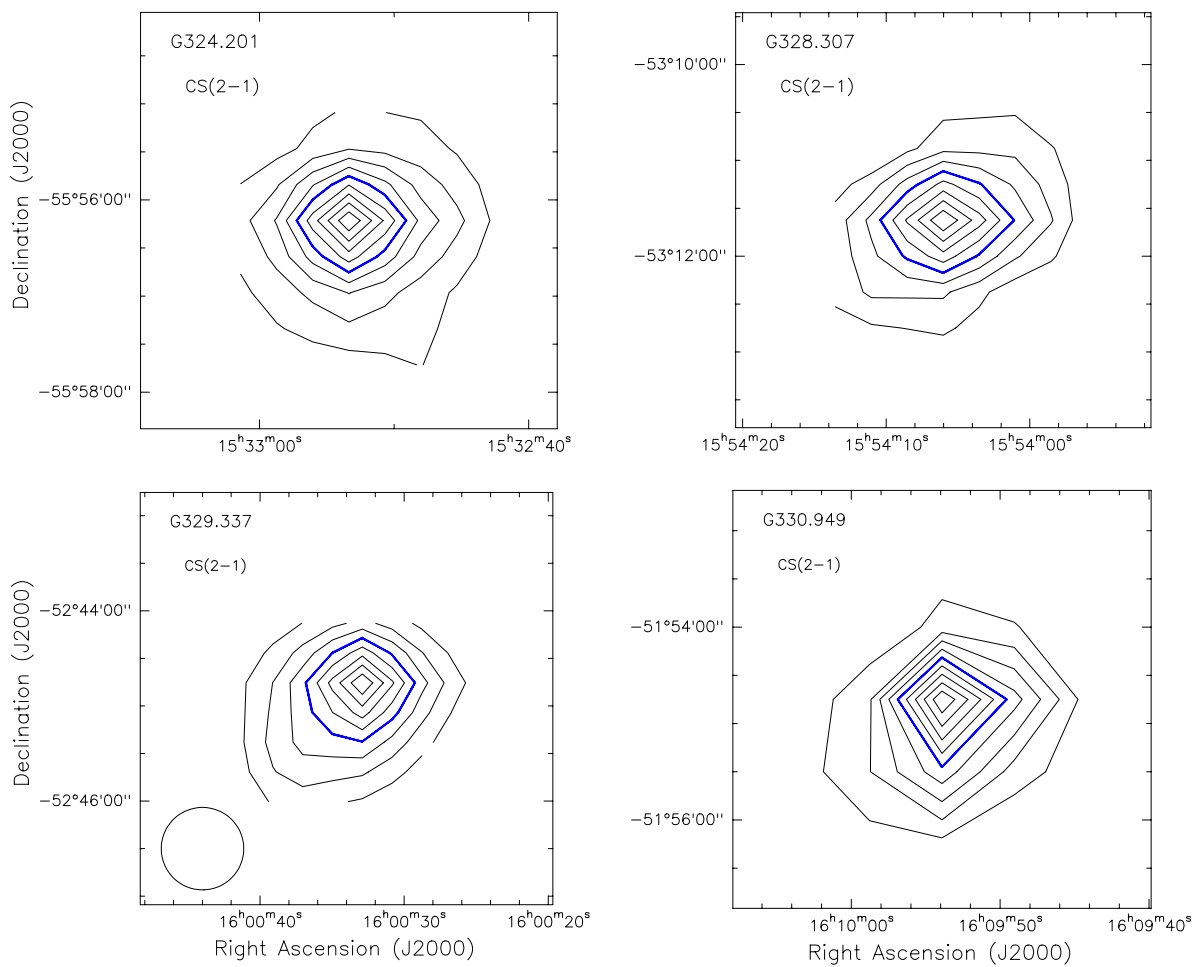}
\caption
{\baselineskip2.0pt
Contour maps of the CS(2$\rightarrow$1) velocity integrated ambient gas emission. 
Contour levels are drawn in steps of 10\% and up to 90\% of the peak 
value. The FWHM beam is shown in the bottom left corner of the lower left panel.   
Upper left: G324.201+0.119.  Peak: 20.8 K~\kms. 
Upper right: G328.307+0.423. Peak: 25.6 K~\kms. Starting contour level: 20\%.  
Lower left: G329.337+0.147. Peak: 28.0 K~\kms. Starting contour level: 20\%. 
Lower right: G330.949-0.174. Peak:  39.7 K~\kms. 
\label{fig-csmaps}}
\end{figure}

Fig.~\ref{fig-csmaps} shows contour maps of the velocity integrated 
ambient gas emission in the CS(2$\rightarrow$1) line toward all four 
regions. The 50\% contour level of the emission is marked with dark lines. 
In all cases the emission, at the resolution of 52\arcsec,  arises form a 
single source exhibiting a simple morphology. 
The deconvolved major and minor FWHM angular sizes determined
by fitting a Gaussian profile to the emission are given in 
\S 3.3.

\subsection{Dust continuum}

\begin{figure}
\epsscale{0.90}
\plotone{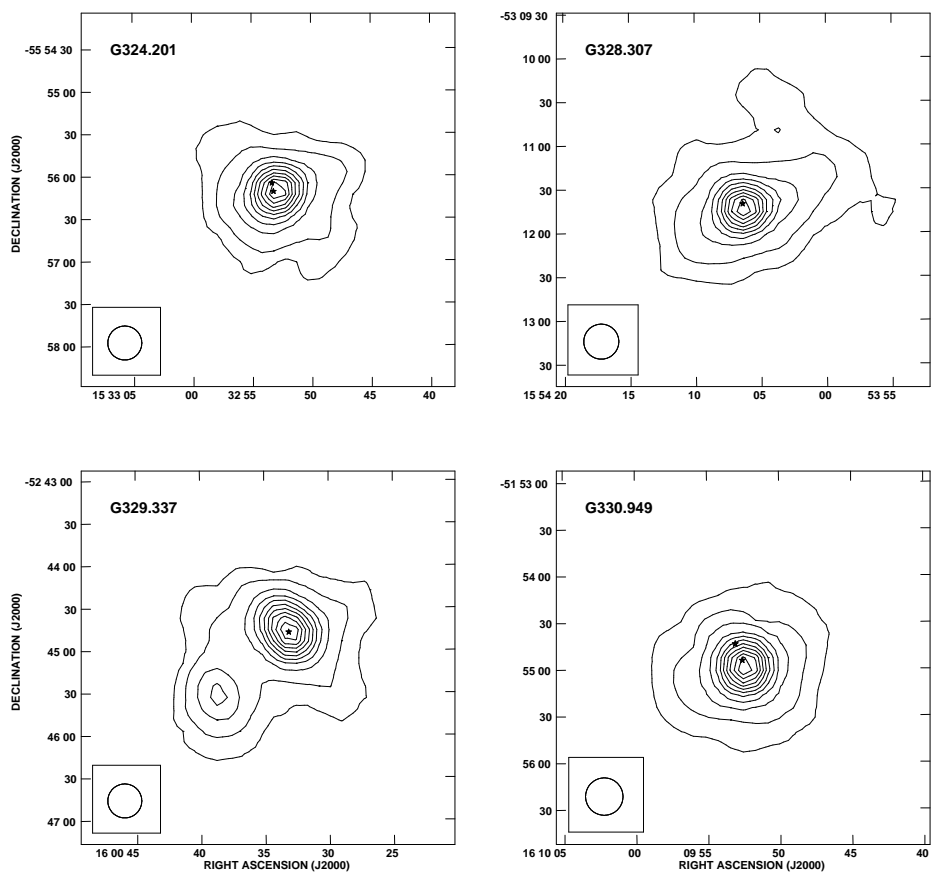}
\caption
{\baselineskip2.0pt
SEST/SIMBA maps of the 1.2-mm dust continuum emission.
Contour levels are drawn at 5, 10, 20, 30, 40, 50 60, 70, 80 and 90\% of the peak
flux density. The FWHM beam is shown at the bottom left.
In all panels, except the upper right, the stars mark the positions of the 
compact radio continuum sources detected by Urquhart et al. (2007). 
Upper left:  G324.201+0.119. Peak: 5.28 Jy beam$^{-1}$.
Upper right: G328.307+0.423. Peak: 6.35 Jy beam$^{-1}$.
The star marks the position of the bright compact radio 
continuum source detected by Garay et al. (2006). 
Lower left:  G329.337+0.147. Peak: 4.35 Jy beam$^{-1}$. 
Lower right: G330.949-0.174. Peak: 21.8 Jy beam$^{-1}$.
\label{fig-dustsimba}}
\end{figure}

Fig.~\ref{fig-dustsimba} presents maps of the 1.2-mm continuum emission,
within regions of typically $4'$ in size, towards the four observed sources. 
Emission was detected toward all four regions. 
The extent and morphology of the 1.2-mm dust continuum emission is similar
to that of the CS\jj{2}{1} line emission, indicating that these two probes 
trace the same physical conditions.
The observed parameters of the 1.2-mm sources are given in
Table~\ref{tbl-obspardust}. Cols.(2) and (3) give their peak position. Cols.(4)
and (5) give, respectively, the peak flux density and the total flux density,
the later measured directly from the maps using the AIPS tasks IMEAN.
Col.(6) gives the deconvolved major and minor FWHM angular sizes determined
by fitting a single Gaussian profile to the whole spatial 
distribution, except for G329.337 for which two Gaussian components were
fitted.

In three of the four regions the morphology of the dust emission 
can be described as arising from a bright compact peak surrounded 
by an extended envelope of weaker emission, which is characteristic
of sources with centrally condensed density profiles. This is illustrated in 
Fig.~\ref{fig-intcuts} which shows slices of the observed 1.2-mm intensity across 
the two sources with nearly circular morphology (G324.201 and G330.949). We 
find that the observed radial intensity profiles can be well fitted with single 
power-law intensity profiles of the form $I \propto r^{-\alpha}$, where $r$ is 
the distance from the center, properly convolved with the beam of 24\arcsec. 
We derive $\alpha$ indices of 1.6 and 1.9 for the G324.201 and G330.949
cores, respectively. For the G328.307 core an $\alpha$ index of 1.5 has been 
already derived by Garay et al. (2007).

\subsection{Individual sources}

Three of the high-mass star forming regions investigated here, G324.201, 
G329.337 and G330.949, are in the list of $\sim$2000 massive young stellar 
objects (MYSO) candidates compiled by the RMS survey based on a colour selection 
criteria of MSX point sources (Hoare et al. 2005). Radio continuum 
observations have, however, shown that all four Norma cores are associated 
with compact radio sources, indicating that they are in a more advanced stage 
of evolution in which an UCHII region has already developed (Walsh et al. 1998; 
Garay et al. 2006; Urquhart et al. 2007). In all cases the compact radio sources 
are found projected at the peak position of the 1.2-mm dust continuum emission, 
suggesting that massive stars are located at the center of the massive and dense 
cores. In what follows we discuss the characteristics of the molecular gas and 
dust emission, as well as of the radio emission, toward each of the IRAS sources, 
individually.

\begin{figure}
\epsscale{0.90}
\plotone{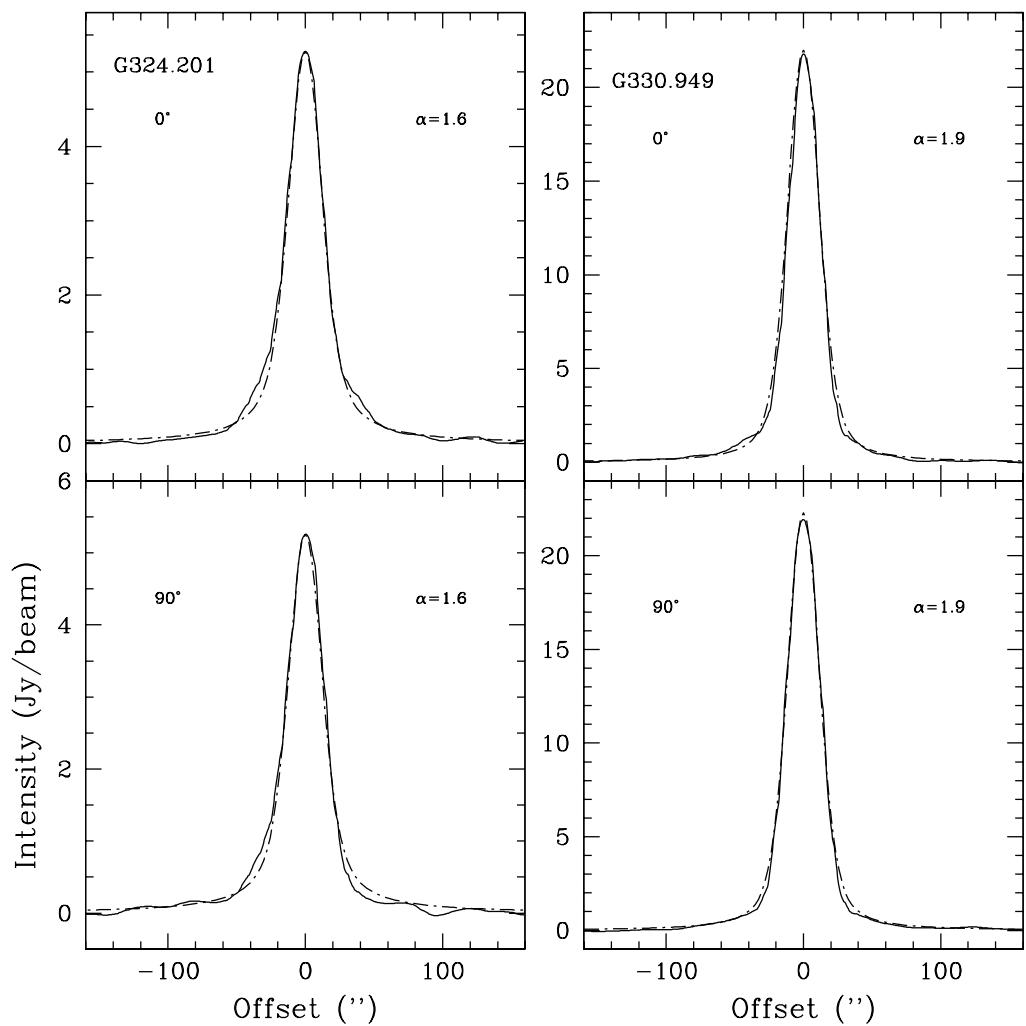}
\caption
{\baselineskip2.0pt
Intensity cuts of the 1.2-mm emission across two orthogonal directions
passing through the peak position of G324.201+0.119 (left panels) 
and G330.949-0.174 (right panels). The position angle of the slices are 
given in the upper left corner of each panel. Dotted lines correspond to 
fits of the observed radial intensity with single power-law intensity profiles. 
The fitted power-law indices are given in the upper right corner of each panel.
\label{fig-intcuts}}
\end{figure}

\noindent
{\sl G324.201+0.119}.$-$
The G324.201+0.119 massive core is associated with the IRAS source 15290-5546, 
which has a total luminosity of $5.9\times10^5 {\cal L}_{\odot}$. The CS profiles 
across the core show emission from a nearly Gaussian component, with central 
velocities of $\sim -88.4$ \kms and FWHM widths of $\sim4.0$\kms. 
The spatial distribution of the emission in the CS($2\rightarrow1$) 
line from the ambient gas is slightly elongated, with FWHM 
deconvolved major and minor axis of 43\arcsec\ and 30\arcsec, respectively. 
The morphology of the 1.2-mm dust emission also shows the presence of 
a single structure with FWHM deconvolved major and minor axis of 
35\arcsec\ and 28\arcsec, respectively. A power-law fit to the 1.2-mm intensity 
profile gives an index of 1.6 (see Fig.~\ref{fig-intcuts}).
The total flux density of the region at 1.2 mm is 17.3 Jy.

Toward the center of the core the Gaussian component is superposed on a broader 
and weaker component. In the CS($3\rightarrow2$) line the wing emission extends 
up to a flow velocity of $-13.1$ \kms\ in the blue side and 11.4 \kms\ in the red 
side. The flow velocity, $v_{flow}$, is defined as $v_{flow} = v-v_o$, where 
$v_o$ is the ambient cloud velocity. The SiO line profiles are, on the other hand, 
dominated by the emission from the broad component, covering a flow velocity range 
from $-12.6$ to $16.8$ \kms. 

Radio continuum observations towards G324.201+0.119 show the presence of two
compact radio sources: a cometary-like component, with an angular
diameter of $\sim$4", located at the center of the core and an unresolved
component located $\sim$6" north of the former (Urquhart et al. 2007).
If excited by single stars, the rate of UV photons needed to produce the
cometary and unresolved regions of ionized gas implies the presence of
O7 and O8.5 ZAMS stars, respectively.

\noindent
{\sl G328.307+0.423}.$-$
The G328.307+0.423 massive core is associated with the IRAS source 15502-5302, 
which has a total luminosity of $1.1\times10^6 {\cal L}_{\odot}$. 
The spatial distribution of the CS($2\rightarrow1$) emission from the ambient 
gas shows an elongated morphology, with FWHM deconvolved 
major and minor axis of 65\arcsec\ and 37\arcsec, respectively. 
The elongated morphology is more clearly seen in the observed spatial distribution 
of the 1.2-mm emission (see Fig.~\ref{fig-dustsimba}), which has  FWHM deconvolved
major and minor axis of 43\arcsec\ and 26\arcsec, respectively. 
The total flux density of the region at 1.2-mm is 24.1 Jy.

The CO($1\rightarrow0$) spectra observed at the center of the core show a double 
peaked profile, with a strong blueshifted peak at the velocity of $-93.9$ \kms 
and a weaker redshifted peak at the velocity of $-88.5$ \kms. 
On the other hand, the spectra of the optically thin C$^{18}$O($1\rightarrow0$) 
line show a single Gaussian profile with a peak velocity of -92.6 \kms.
These spectral characteristics indicate the presence of infall motions (Leung \& 
Brown 1977, Walker, Narayanan, \& Boss 1994, Myers et al. 1996). Using expression 
(9) of Myers et al. 
(1996) we derive an infall velocity of $\sim 0.3$ \kms. The profiles of the CS and 
C$^{34}$S emission observed across the core are generally asymmetric, showing a 
peak at velocities of $\sim-92.8$ \kms and a shoulder toward redshifted 
velocities. The observed shapes are consistent with the infall motions suggested
by the CO($1\rightarrow0$) emission but for transitions with smaller optical depths. 
In addition to infalling gas, outflowing gas is clearly detected toward the central 
region. In the CS(3$\rightarrow$2) line, blueshifted and redshifted emissions are 
detected up to flow velocities of $-8.2$ \kms\ and 12.2 \kms, respectively. 
The SiO emission, which is detected in the velocity range from $-100.0$ to $-82.6$ 
\kms, arises mainly from the outflow component, but a narrower component from 
the ambient core gas is also present. 

Radio continuum observations towards G328.307+0.423 show the presence of a 
complex region of ionized gas, consisting of a handful of low brightness, extended
components and a bright, compact component located at the center of the
core (Garay et al. 2006).  The flux density of the later component
increases with frequency whereas its angular size decreases with frequency.
These characteristics suggest the presence of a steep gradient in the electron
density.

\noindent
{\sl G329.337+0.147}.$-$ 
The G329.337+0.147 massive core is associated with the IRAS source 15567-5236, 
which has a total luminosity of $1.2\times10^6 {\cal L}_{\odot}$. The CS 
profiles across the core show emission from a bright component with a central 
velocity of $\sim-107.7$ \kms and a FWHM width of $\sim4.5$ \kms. 
Strong high velocity wing 
emission is detected toward the central region in all observed species (see 
Fig.~\ref{fig-specpeakg329}). The blueshifted and redshifted emissions extend up 
to flow velocities of $-27.6$ \kms\ and $24.2$ \kms\ in the CO(1$\rightarrow$0) 
line, $-14.8$ \kms and $16.8$ \kms\ in the CS(3$\rightarrow$2) line, and $-19.2$ 
\kms\ and 17.4 \kms\ in the SiO(3$\rightarrow$2) line.

\begin{figure}
\epsscale{0.90}
\plotone{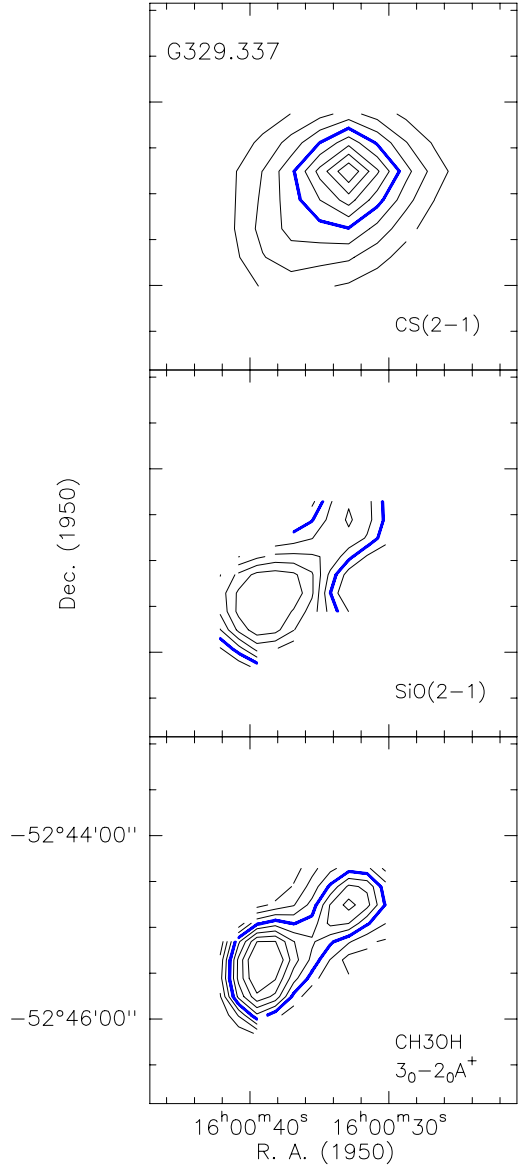}
\caption
{\baselineskip3.0pt
Contour maps of velocity integrated ambient molecular gas emission 
towards G329.337+0.147. 
Contour levels are drawn at 20, 30, 40, 50 60, 70, 80 and 90\% of the peak
value.  The star marks the position of the radio source detected by Walsh et al. 
(1997).
Upper: CS(2$\rightarrow$1). Peak value: 28.0 K~\kms.
Center: SiO(2$\rightarrow$1). Peak value: 1.45 K~\kms. 
Lower: CH$_3$OH $3_0\rightarrow2_0$ A$^+$. Peak value: 3.65 K~\kms.
\label{fig-mapsg329}}
\end{figure}

The CS($2\rightarrow1$) emission from the ambient gas arises from 
a structure that is slightly elongated in the SE direction, with deconvolved FWHM 
major and minor axis of 44\arcsec\ and 39\arcsec, respectively. The peak of the 
CS emission is coincident with a methanol maser (Walsh et al. 1997). The spatial 
distribution of the silicon monoxide 
and methanol emission show notable differences with respect to that of the 
emission in carbon monosulfide.  The morphology of the SiO and 
CH$_3$OH emission shows a double peaked structure, with the main peak 
being displaced from the peak of the CS emission by $\sim$ 60\arcsec\ toward the 
southeast (see Fig.~\ref{fig-mapsg329}). The secondary SiO peak coincides 
with the peak of the CS emission. The double peaked morphology is also clearly 
seen in the map of the dust continuum emission.

Radio continuum observations towards G329.337+0.147 show the presence
of a single source, with an angular diameter of $6.3\times4.8$", exhibiting an
irregular multi-peaked morphology (Walsh et al. 1998; Urquhart et al. 2007). 
If excited by a single star, the rate of UV photons required to ionize this 
region implies the presence of an O6.5 ZAMS star.
 
\noindent
{\sl G330.949-0.174}.$-$
The G330.949-0.174 massive core is associated with the IRAS source 16060-5146,
which has a total luminosity of $1.0\times10^6 {\cal L}_{\odot}$. 
The CS(2$\rightarrow$1) line exhibit an asymmetric 
profile, with a peak at the velocity of $-92.1$ \kms and a shoulder toward 
redshifted velocities. The CS(3$\rightarrow$2) and CS(5$\rightarrow$4) profiles 
are double peaked, with peaks at $\sim-93.0$ \kms and $\sim-89.1$ \kms. The blue 
peak is stronger than the red peak by a factor of $\sim$ 1.5. The profiles of 
the C$^{34}$S and SiO emission are, on the other hand, more symmetric, with peak 
velocities lying between the two velocity peaks of the CS emission. This result 
indicates that the complex CS profiles are produced by optical depth effects 
rather than being due to two velocity components along the line of sight.

The presence of a high velocity gas toward the central region is clearly 
indicated by the profiles of the CS and SiO emission. In CO lines the outflow 
emission is blended with emission from less dense molecular clouds in the line 
of sight.  The emission in the CS($3\rightarrow2$) line extends up to a 
velocity of $-75.7$ \kms in the red side and up to a velocity of $-106.0$ \kms in 
the blue side.

The spatial distribution of the CS($2\rightarrow1$) emission from the ambient 
gas is nearly circular, with FWHM deconvolved major and minor axis of 43\arcsec. 
The circular morphology is also seen in the 1.2-mm dust emission,
which exhibits a single structure with FWHM deconvolved major and minor axis of
33\arcsec. The total flux density of the region at 1.2 mm is 47.2 Jy.
A power-law fit to the 1.2-mm intensity profile gives an index of $1.9\pm0.1$
(see Fig.~\ref{fig-intcuts}).
The peak of the molecular and dust emissions are displaced about 30\arcsec\ east 
of the IRAS position, but are coincident with the position of a methanol maser and a 
compact radio continuum source (Walsh et al. 1997, 1998).
The radio continuum observations of Urquhart et al. (2007) towards G330.949-0.174 
show the presence of two compact sources: a bright cometary-like component, with 
an angular diameter of $\sim$3", located near the center of the core and an 
unresolved, weak component, located $\sim$12" northeast of the former. If excited 
by single stars, the rate of UV photons needed to produce the cometary and 
unresolved regions of ionized gas implies the presence of O7 and B0 ZAMS stars, 
respectively.

\vfill\eject 

\section{DISCUSSION}

\subsection{Spectral Energy Distributions}

\begin{figure}
\epsscale{0.95}
\plotone{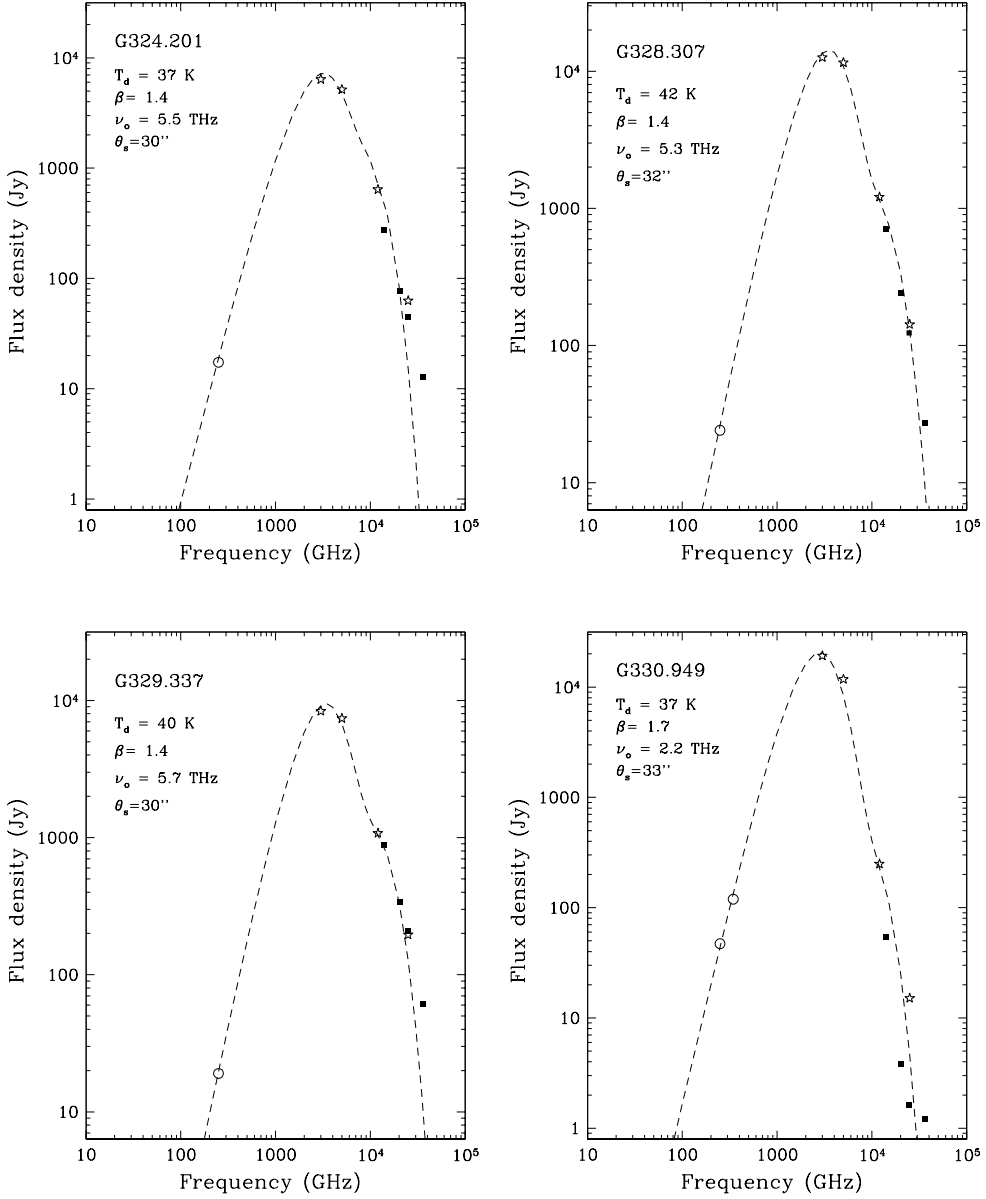}
\caption
{\baselineskip3.0pt
Spectral energy distributions.
Stars mark IRAS fluxes, squares MSX fluxes, and the circle the SIMBA flux.
The long-dashed curve is a fit to the spectrum using two modified blackbody functions
of the form $B_{\nu}(T_d) \left[ 1 - \exp(-(\nu/\nu_o)^{\beta}) \right]$, with
different temperatures. The short-dashed line indicates the fit for the colder
temperature component (fit parameters indicated on the upper left).
\label{fig-sed}}
\end{figure}

Fig.~\ref{fig-sed} shows, for each of the 4 sources, the spectral energy 
distribution (SED) from 12 $\mu$m to 1.2 mm. In this wavelength range the 
emission is mainly due to thermal dust emission.  The flux densities at 12, 25, 
60 and 100~\micron\ correspond to IRAS data, and those at 8.3, 12.1, 14.7 and 
21.3~\micron\ correspond to \emph{Midcourse Space Experiment (MSX)} data 
(Price et al. 2001). The IRAS and MSX fluxes were obtained from
the IPAC database.

We fitted the SED with modified blackbody functions of the form 
$B_{\nu}(T_d)\left[1-\exp(-\tau_{\nu})\right]\Omega_s~, $
where $\tau_{\nu}$ is the dust optical depth, $B_{\nu}(T_d)$ is the Planck
function at the dust temperature $T_d$, and $\Omega_s$ is the solid angle subtended
by the dust emitting region. The opacity was assumed to vary with frequency as
$\nu^{\beta}$, i.e. $\tau_{\nu}= \left(\nu/\nu_o\right)^{\beta}$, where
$\nu_o$ is the frequency at which the optical depth is unity. A single temperature 
model produced poor fits, underestimating the emission observed at wavelengths 
smaller than 25$\mu$m, and therefore we used a model with two temperature 
components. Since the regions are likely to present temperature gradients, this 
is indeed a coarse simplification. However, this model allows us to determine 
the average dust temperatures representative of each wavelength range.
The parameters derived from the fits, namely dust temperature, 
opacity power-law index, wavelength at which the opacity is unity, and 
angular size (assuming a Gaussian flux distribution), for the colder dust 
component are given in Fig.~\ref{fig-sed}. 
  
\subsection{Parameters derived from dust observations}

The parameters derived from the 1.2-mm observations are summarized in
Table~\ref{tbl-deriveddust}. The dust temperatures given in col. (2) correspond 
to the colder temperature determined from fits to the spectral energy 
distribution (SED). The radius given in col. (3) were computed from the geometric 
mean of the deconvolved major and minor angular sizes obtained from single 
Gaussian fits to the observed spatial structure. The masses of the cores, given 
in col.(4), were derived using the expression, (e.g., Chini, Kr\"ugel, 
\& Wargau 1987)
\begin{equation}
M_{g} = {{S_{1.2mm} D^2}\over{R_{dg} \kappa_{1.2mm} B_{1.2mm}(T_d)}} ~~, 
\label{eqn-mdust}
\end{equation}
where $S_{1.2mm}$ is the observed flux density at 1.2mm, $\kappa_{1.2mm}$ is the 
mass absorption coefficient of dust with a value of 1 cm$^2$ g$^{-1}$ (Ossenkopf 
\& Henning 1994), $R_{dg}$ is the dust-to-gas mass ratio (assuming 10\% He) 
with a value of $0.01$, and $B(T_d)$ is the Planck function at the dust temperature 
$T_d$. This expression implicitly assumes that the observed emission at 1.2 mm
corresponds to optically thin thermal dust emission and that the source is 
isothermal. The masses derived in this way range from $5\times10^3$ to $1\times10^4$ 
\mo.  Cols.~(5) and (6) give, respectively, the average molecular densities
and average column densities derived from the masses and radius assuming that 
the cores have 
uniform densities. Clearly, this is a rough simplification, since as discussed in 
\S 3.2 the massive and dense cores are likely to have steep density gradients. 
Finally, the continuum optical depth at 1.2-mm is given in col.~(7).

\subsection{Parameters derived from line observations}

\subsubsection{Optical depths and column densities of CS and CO}

The optical depth in the \cstd\ line, $\tau_{32}^{\small CS}$, can be derived from 
the ratio of observed main beam brightness temperatures in the \cstd\ and 
\ctcstd\ lines using the expression, 
\begin{equation}
{{1-\exp(-\tau_{32}^{\small CS}/r)}\over{1-\exp(-\tau_{32}^{\small CS})}} =
{{[J_{32}^{\small CS}(T_{ex})-J_{32}^{\small CS}(T_{bg})]} 
 \over{[J_{32}^{\small C^{34}S} (T_{ex})-J_{32}^{\small C^{34}S} (T_{bg})]}} 
 {{T_{mb}(C^{34}S)}\over{T_{mb}(CS)}} ,~
\label{eqn-optdepth}
\end{equation}
where the subscript 32 refers to the $3\rightarrow2$ transition, 
$T_{ex}$ is the excitation temperature of the transition, $T_{bg}$ 
is the background temperature, 
\begin{displaymath}
J_{\nu}(T) = \frac{h\nu}{k}\frac{1}{\exp(h\nu/kT)-1} ~~~ ,
\end{displaymath}
and $r$ is the $\tau_{32}^{\small CS}/\tau_{32}^{\small C^{34}S}$ optical depth 
ratio, given by
\begin{equation}
r = a {{\mid\mu^{CS}\mid^2}\over{\mid\mu^{C^{34}S}\mid^2}}
 \left\{{{\exp(-12 h B^{\small CS}/k T_{ex})}\over{\exp(-12 h 
 B^{\small C^{34}S}/k T_{ex})}}\right\}
\left\{{{\exp(h\nu_{32}^{\small CS}/kT_{ex})-1}\over
{\exp(h\nu_{32}^{\small C^{34}S}/kT_{ex})-1}}\right\}
 {{(k T_{ex}/h B^{\small C^{34}S} + 1/3)}\over{(k T_{ex}/h B^{\small 
 CS} +1/3)}} ~~, 
\label{eqn-TauRatio}
\end{equation}
where $a$ is the [C$^{32}$S/C$^{34}$S] isotopic abundance ratio, and $\mu$ and $B$ 
are the permanent dipole moment and the rotational constant of the molecule, 
respectively. 
If the excitation temperature is known, the above expressions allow the optical 
depth to be determined. We avoid the usual approximation that $\tau_{32} \gg 1$, 
and solve equation~(\ref{eqn-optdepth}) using an interpolation procedure,
assuming a [C$^{32}$S/C$^{34}$S] abundance ratio of 22.5 (Blake et al. 1994).  

The optical depth in the CO($1\rightarrow0$) line, $\tau_{10}^{\small CO}$, can be 
derived from the ratio of observed main beam brightness temperatures in the 
CO($1\rightarrow0$) and $^{13}$CO($1\rightarrow0$) lines in a similar manner as 
described above (see Bourke et al. 1997, for the appropriate expressions).
We assumed a [CO/$^{13}$CO] abundance ratio of 55, corresponding
to the average value across the Galactic disk (Wannier et al. 1982).
Further, the observations of the J=$1\rightarrow0$ line emission in the three 
isotopic species of carbon monoxide (CO, $^{13}$CO and C$^{18}$O) 
allow to determine the [$^{13}$CO]/[C$^{18}$O] isotopic abundance ratio.  
Using a [CO]/[$^{13}$CO] abundance ratio of 55, we derive 
[$^{13}$CO]/[C$^{18}$O] abundance ratios in the range 8 to 10.

The peak optical depths at the peak position of the massive cores in the \cstd, 
\ctcstd, \couc\ and \trecouc\ lines, are given in cols. 2 to 5 of 
Table~\ref{tbl-opacities}. They were computed assuming $T_{ex}$ = 40 K. We 
find that the emission is optically thick in the CS lines and optically thin in 
the C$^{34}$S line for all cores.
We note that the dependence of the optical depth with excitation temperature is 
weak. For instance, assuming $T_{ex}$ = 100 K the derived optical depths are 
only 0.4\% larger. 

The total column density $N$ of a linear, rigid rotor, molecule can be derived 
from the optical depth and excitation temperature 
of a rotational transition at frequency $\nu$, using the expression 
(e.g. Garden et al. 1991) 
\begin{equation}
N = \frac{3h}{8\pi^3\mu^2} \frac{k(T_{ex}+hB/3k)}{(J+1)hB}
\frac{\exp(E_J/kT_{ex})}{1 - \exp(-h\nu/kT_{ex})} \int \tau dv ~~, 
\label{eqn-Ntot}
\end{equation}
where $J$ is the rotational quantum number of the lower state. This expression 
assumes that all energy levels are populated according to local thermodynamic 
equilibrium (LTE) at the temperature $T_{ex}$. In particular the total column 
density of the CS molecule is given, in terms of the opacity and excitation 
temperature of the $3\rightarrow2$ transition, by 
\begin{equation}
N({\rm CS}) =  5.93\times10^{11}~ (T_{ex}+0.39)
\frac{\exp(7.05/T_{ex})}{1-\exp(-7.05/T_{ex})} 
\int \tau_{32}^{\small CS}~ dv ~~, 
\label{eqn-nc34s}
\end{equation}
where $v$ is measured in \kms. Col. 2 of Table~\ref{tbl-colden} gives 
the derived CS column density through the center of the cores.
The CO column density through the center of the cores are given in 
col. 3 of Table~\ref{tbl-colden}. They were determined in the same way 
as the CS column densities using the [CO,$^{13}$CO] pair of observations.

From the column density in CS, N(CS), we can estimate the mass of the cores using 
the expression,
\begin{equation}
 M = N(CS) \left[{{H_2}\over{CS}}\right] \mu_m~ m_H~ \pi R^2~~,
\end{equation}
where $[{{H_2}\over{CS}}]$ is the H$_2$ to CS abundance ratio, $\mu_m$ is the mean
molecular weight, and $R$ is the core radius. The radii, given in column 2
of Table~\ref{tbl-derivedpar}, were computed from the geometric mean of the
deconvolved CS(2$\rightarrow$1) angular sizes using the distances given in
Table~\ref{tbl-obssources}.
The core masses computed from expression (6), using $[{{H_2}\over{CS}}]=
3\times10^7$ (see \S 4.3), are given in col. 3 of Table~\ref{tbl-derivedpar}.
They range from $4.3\times10^3$ to $1.0\times10^4$ \mo.
These masses are in good agreement with those
derived from the dust continuum observations. 

\subsubsection{Rotational temperatures and column densities of methanol
and silicon monoxide}

The rotational temperature, $T_{rot}$, and the total column density, $N_T$, of 
methanol and silicon monoxide can be computed using a rotational diagram analysis 
(see e.g., Linke, Frerking \& Thaddeus 1978; Blake et al. 1987), with the 
assumptions of optically thin conditions and local thermodynamical equilibrium 
(LTE), which relates the integrated line intensity, rotational temperature, and 
column density via
\begin{equation}
\frac{3 k \int T_{mb} dv}{8 \pi^3 \mu^2 \nu S} = \frac{N_u}{g_u}
 = \frac{N_T}{Q(T_{rot})} \exp \left(-\frac{E_u}{kT_{rot}}\right)~~,
\end{equation}
where $\mu$, $\nu$, and $S$ are the transition dipole moment, frequency, 
and line strength of the transition, respectively, $\int T_{mb} dv$ is the 
velocity integrated main beam brightness, obtained directly from the 
observations, $E_u$ is the upper state energy, and $Q(T_{rot})$ is the 
rotational partition function.  

\begin{figure}
\epsscale{0.95}
\plotone{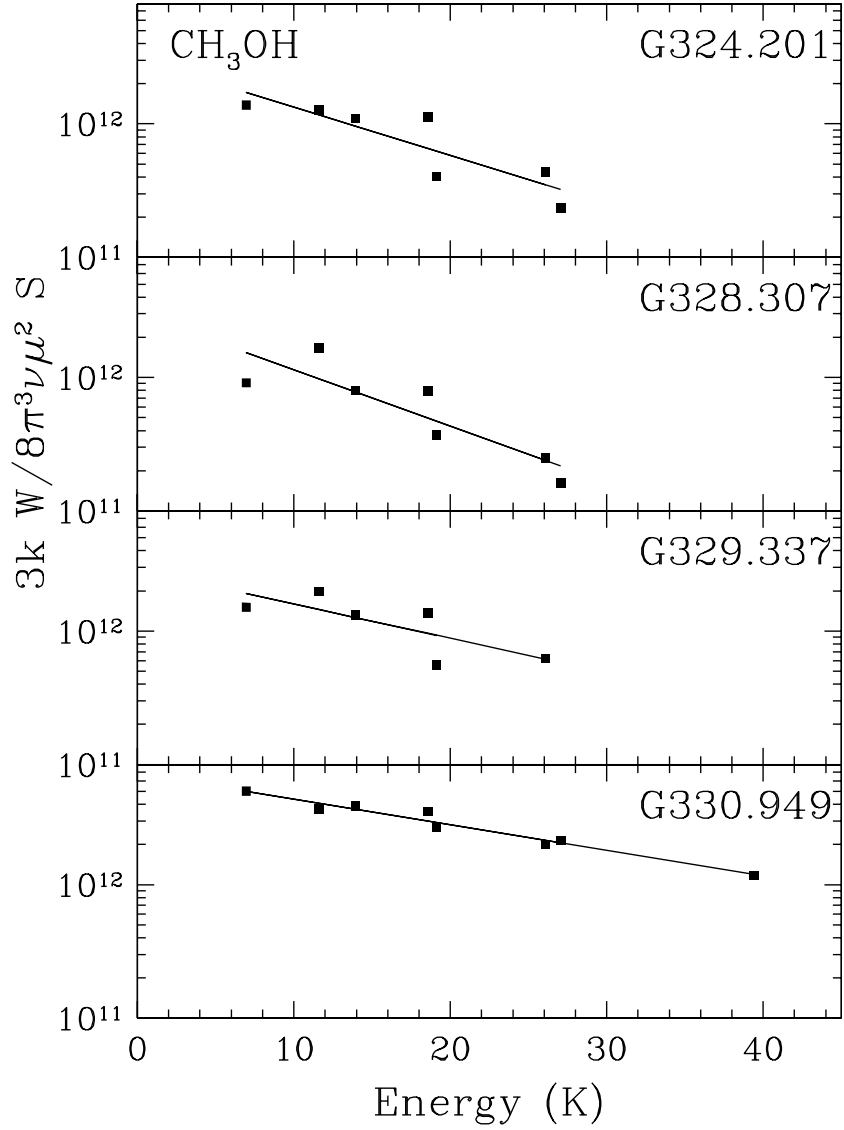}
\caption
{\baselineskip3.0pt
Rotational diagrams for the methanol emission observed toward
the massive cores. The lines correspond to least squares linear
fits to the observed data. The derived values of the rotational
temperature and total column density are given in Table~\ref{tbl-abundances}. 
\label{fig-trotmet}}
\end{figure}

Fig.~\ref{fig-trotmet} shows rotational diagrams for the CH$_3$OH emission 
observed at the peak position of the massive cores. The column densities and 
rotational temperatures, derived from a linear least squares fit to the data, are 
given in cols. 4 and 6 of Table~\ref{tbl-colden}. The rotational temperatures 
of methanol range from 12.0 K to 22.6 K. These temperatures are significantly 
smaller than the temperatures derived from the dust observations, of $\sim$ 40K,
indicating that the methanol populations are sub-thermally excited 
(Bachiller et al. 1995; Avery \& Chiao 1996). 

\begin{figure}
\epsscale{0.85}
\plotone{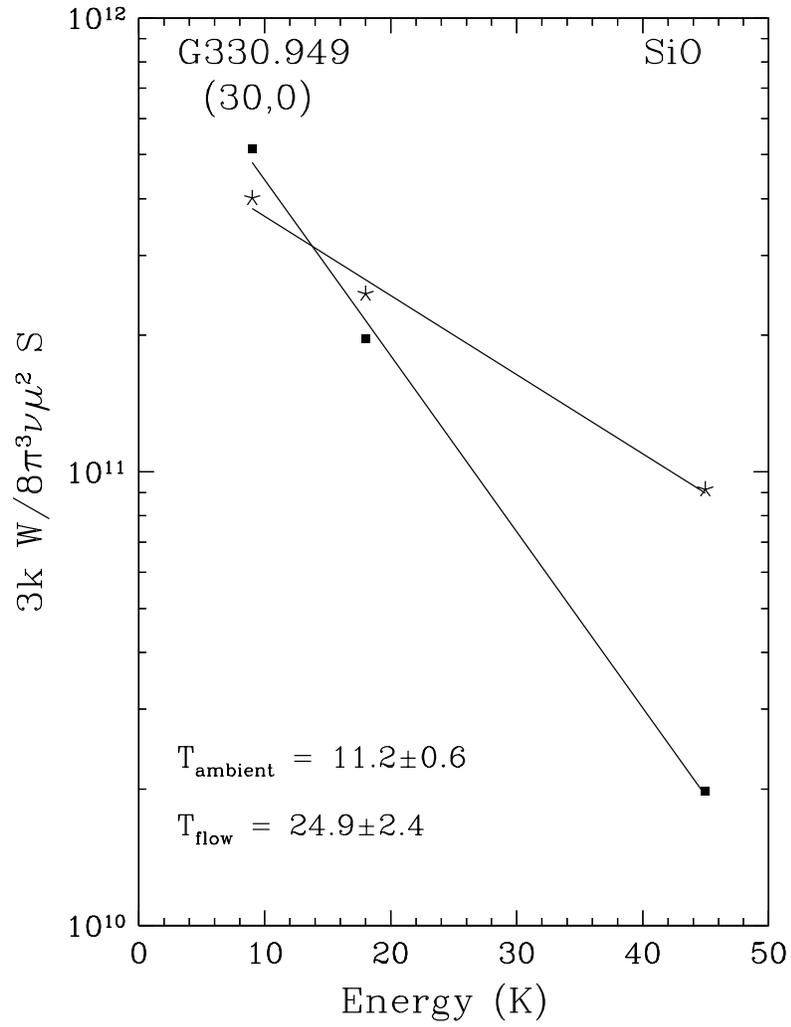}
\caption
{\baselineskip3.0pt
Rotational diagram for the narrow (squares) and broad (stars) SiO emission 
observed towards the G330.949-0.174 massive core. The lines correspond to least 
squares linear fits to the observed data. The derived values of the rotational
temperature are given in the lower left corner.
\label{fig-trotsio}}
\end{figure}

From rotational diagram analysis of the SiO emission observed at the peak position 
of the massive cores we derived column densities and rotational temperatures 
of both the ambient gas, given in cols. 5 and 7 of Table~\ref{tbl-colden}, and 
flowing molecular gas, given in 
cols. 8 and 9.  The rotational temperatures of the flowing gas are typically
two times larger than those of the ambient gas. This is illustrated in
Fig.~\ref{fig-trotsio}, which presents a rotational diagram for the SiO emission
observed at the peak position of the G330.949 massive core. In this case the gas
giving rise to the narrow emission has a rotational temperature of 11.2$\pm0.6$ K
and a column density of $4.1\times10^{13}$ cm$^{-2}$, whereas the outflowing gas
producing the broad emission has a rotational temperature of 24.9$\pm2.4$ K and
a column density of $2.6\times10^{13}$ cm$^{-2}$.

\subsubsection{Molecular abundances}

The chemical composition of a gas cloud is usually characterized by the fractional 
abundance of molecules relative to molecular hydrogen, the main constituent of 
the interstellar gas. The abundance of H$_2$ is however not directly measured. 
Our observations allow us, however, to directly determine abundances 
relative to CO. We compute the abundance of species X relative to CO, [X/CO] 
(where X = CS, SiO, or CH$_3$OH), as the ratio of the column density of species 
X and the CO column density in the same velocity range.
For SiO and CH$_3$OH we use the column densities 
derived from rotational analysis (see \S x.x) whereas for CS and CO we 
use the column densities derived from the observations of their main and isotopic 
species (see \S 4.3.1). Table~\ref{tbl-abundances} summarizes
the abundances of CS, SiO and CH$_3$OH relative to CO. 
To convert [X/CO] to fractional abundances relative 
to H$_2$ we will assume [CO/H$_2$]=$10^{-4}$. 

\noindent 
{\bf Silicon monoxide}. We find that the [SiO/CO] abundance ratio in 
the ambient gas range from $3\times10^{-7}$ to $1\times10^{-6}$. Assuming that 
[CO/H$_2$] is $10^{-4}$, then the [SiO/H$_2$] abundance ratio in the ambient gas 
ranges between $3\times10^{-11}$ and $1\times10^{-10}$. Chemical models predict 
that both high temperature and shocks can produce enhancements in the gas phase 
SiO abundance. The abundance of SiO in the ambient gas is likely due to the 
gas-phase high temperature chemistry that takes place in the hot regions surrounding 
the recently formed massive stars. Models of gas-phase chemistry of hot molecular 
cores that takes into account grain mantle evaporation by stellar and/or shock 
radiation (MacKay 1995), predict [SiO/H$_2$] abundances of $3\times10^{-11}$ 
for gas kinetic temperatures of 100 K and densities of $2\times10^{6}$ cm$^{-3}$.   

We also find that the [SiO/CO] abundance ratios in the outflowing gas are 
similar to those derived in the ambient gas. The abundance of SiO in the high 
velocity gas is likely to be due to gas phase chemistry behind a shock 
(Iglesias \& Silk 1978; Hartquist et al. 1980; Mitchell \& 
Deveau 1983). The enhancement of the SiO abundance in the high velocity material 
is probably the result of grain destruction by shocks which releases silicon 
into the gas phase, allowing SiO to form (Seab \& Shull 1983; Schilke et al. 1997; 
Caselli et al. 1997).

\noindent 
{\bf Methanol}. We find that the [CH$_3$OH/CO] abundance ratio in the 
ambient gas is similar for all observed cores, having a mean value of 
$6\times10^{-5}$. Assuming an [H$_2$/CO] 
of $10^4$, it implies that the [CH$_3$OH/H$_2$] abundance ratio in the ambient 
gas is typically $6\times10^{-9}$. At temperatures smaller than 100 K, chemistry 
models predict that the production of CH$_3$OH in the gas phase yield abundances 
of only 10$^{-11}$ relative to H$_2$ (Lee et al. 1996). The high abundance of 
methanol in the cores suggests that the radiation from 
their embedded luminous stars are producing a substantial  
evaporation of icy grain mantles. Van der Tak et 
al. (2000a) found a [CH$_3$OH/H$_2$] ratio of a few $10^{-8}$ for hot cores and 
$\sim10^{-9}$ for the cores with the lower temperatures. The values obtained 
for the cores investigated here are in between the values 
found for dark clouds and hot cores, and may correspond to an average 
of the values appropriate for the cold envelope mass and hot core mass.

\noindent 
{\bf Carbon monosulfide}. We find that the [CS/CO] abundance ratio
in the ambient gas is similar for all observed cores, having a mean value of
$4\times10^{-4}$. Assuming an [H$_2$/CO] of $10^4$, it implies that the
[H$_2$/CS] abundance ratio in the ambient gas is typically $3\times10^{7}$.
This value is typical of molecular cores associated with UC HII
regions and hot cores (van der Tak et al. 2000b).

\subsubsection{Structure of the cores}

In the previous analysis we have implicitly assumed that the density and 
temperature are uniform within the massive cores. This is
obviously a simplification, since the physical conditions are likely to vary
with distance from the center of the core. In fact, the intensity profiles 
of the dust continuum emission indicate that the cores are highly centrally 
condensed.

Assuming that cores have density and temperature radial distributions
following power laws, then for optically thin dust emission the intensity index
${\alpha}$ is related to the density index $p$ ($n \propto r^{-p}$) and the
temperature index $q$ ($T \propto r^{-q}$), by the expression
$\alpha = p + Qq -1 + \epsilon_f $ (Adams 1991; Motte \& Andr\'e 2001), where Q 
is a temperature and frequency correction factor with a value of
$\sim1.2$ at 1.2-mm and 30 K (Beuther et al. 2002) and $\epsilon_f$ 
is a correction factor to take into account the finite size of the cores. The 
massive and dense
cores investigated here are heated by a cluster of luminous sources
located at their central position (Chavarr\'\i a et al. 2009; Paper II).
For this type of objects van der Tak et al. (2000b) found that the temperature 
decreases with distance following a power-law with an index of 0.4. Adopting this 
value for $q$ and a value of 0.1 for $\epsilon_f$, we then infer, using the above 
expression, that massive and dense cores have density distributions with 
power-law indices in the range $1.9 - 2.3$.
Several dust continuum studies have already shown the presence of density
gradients within massive and dense cores (e.g., van der Tak et al. 2000b,
Mueller et al. 2002, Beuther et al. 2002, Williams et al. 2005).
In particular, Beuther et al. (2002) observed a large sample of high-mass 
protostellar candidates and found that the steeper indices 
are associated with the more luminous and more massive objects.
The steep indices derived for the Norma cores are characteristics of 
the subgroup of {\sl strong molecular sources} within Beuther's sample, which 
have a mean density index of 1.9.  The steep profiles are thought to indicate
objects in the collapse and accretion phase.

\subsection{Segregation within the massive and dense cores}

Given the large mass of the massive and dense cores investigated here, of 
typically $5\times10^3$ M$_\odot$, their overall collapse process is likely 
to produce a protostar cluster. The questions of how an individual massive star 
forms and how the cluster forms are therefore closely related. The derived 
properties of massive and dense cores are then important boundary conditions to 
be taken into account for models and simulations. 

For dynamical considerations, the relevant mass of the cores is their total mass,
including gas and stars.  We have already shown that the mass of gas in the cores 
determined from two independent methods, one based on dust observations and the 
other on molecular line observations, are in very good agreement. The 
determination of the total mass of the cores is not straightforward and is usually 
estimated assuming that the cores are in virial equilibrium. Masses computed 
under this assumption, using the observed velocity dispersion and size of the 
CS(2$\rightarrow$1) emission, 
are given in column 4 of Table~\ref{tbl-derivedpar}. The virial masses are 
similar, within the errors, to the gas masses, implying that the bulk of the core 
mass is in the form of molecular gas. We note that the virial masses are in 
average a factor of 1.5 smaller than the dust or molecular masses which might also 
indicate that the cores do not have enough support and may be contracting. 

In paper II we show that the massive cores investigated here are associated 
with clusters of embedded stars which exhibit clear mass segregation. 
The high-mass stars are found concentrated near the center of the core while 
intermediate mass stars are spread over much larger distances from the 
core center. Garay et al. (2006) have already reported that massive 
stars are typically located near the center of the massive and dense cores. 
The question arises as to whether the massive stars are formed in situ or if 
their spatial location is a result of dynamical effects.

The characteristics of the cores derived here, namely their high densities, 
steep density profiles and sizes, together with the relatively low fraction of 
the total mass in the form of stars, suggest that dynamical friction (also 
called gravitational drag) produced by the gaseous background onto the stars 
may play an important role in the orbital evolution of stars (Ostriker 1999,
Sanchez-Salcedo \& Brandenburg 2001, Kim \& Kim 2007). The role of gravitational 
drag on a gaseous background in the evolution and migration of stars within dense 
gaseous star forming clouds has been investigated using numerical smooth particle 
hydrodynamics simulations (Escala et al. 2003, 2004). Simulations have been 
carried out with initial conditions similar to those of the dense cores investigated 
here, namely steep cloud density profile ($n(r) \propto r^{-1.8}$) and perturber 
having masses of 1~\% of the total gas mass ($\rm M_{star}\sim20$\Msun). 
For a typical cloud mass of $4\times10^{3}$~\Msun\ and initial distance of the 
stars to the center of the cloud of 0.4~pc, the migration timescale for high-mass
stars is $4.5\times10^{5}$~yr. This value is smaller than the estimated ages 
of the young clusters within the cores of $\sim$ 1~Myr (see paper II).
Since the migration timescales for dynamical friction are proportional to 
the inverse of the perturber's mass (Binney \& Tremaine 1987), intermediate mass 
stars have migration 
timescales typically a factor $\sim$ 10 longer than those of high mass stars 
and greater than the estimated age of the clusters. Therefore, 
gravitational drag is considerably less effective for intermediate mass stars 
than for high mass stars. This suggest that in the young embedded stellar clusters 
only the massive stars have been able to efficiently migrate towards the center 
due to gravitational drag. 
We conclude that dynamical interactions of cluster members with the gaseous 
component of massive and dense cores can explain the observed mass segregation, 
indicating that the gas play an important role in the dynamic of young cluster 
members and originating the observed mass segregation.

We can not rule out, however, the hypothesis that the cradles of massive stars
are formed by direct accretion at the center of centrally condensed massive and
dense cores (McKee \& Tan 2003). In their turbulent and pressurized dense core 
accretion model, the collapse of a massive and dense core is likely to produce 
the birth of a stellar cluster, with most of the mass going into relatively 
low-mass stars. The high-mass stars are formed preferentially at the center of 
the core, where the pressure is the highest, and in short time scales of 
$\sim10^5$ yrs (Osorio et al. 1999; McKee \& Tan 2002).

\section{SUMMARY}

We undertook molecular line observations in several species and 
transitions and dust continuum observations, both made using SEST,
towards four young high-mass star forming regions associated with 
highly luminous IRAS point sources with colors of UC H{\small II} regions and 
CS(2$\rightarrow$1) emission. These are thought to be massive star forming 
regions in early stages of evolution. The objectives were to determine the 
characteristics and physical properties of the dust clouds in which high-mass 
stars form.  Our main results and conclusions are summarized as follows.

We find that the luminous star forming regions in Norma are associated
with molecular gas and dust cores with radii of typically 0.5 pc,
masses of $\sim 5\times10^3$ M$_\odot$, molecular hydrogen densities of 
typically $\sim2\times10^{5}$ cm$^{-3}$, column densities of 
$\sim 5\times10^{23}$ cm$^{-2}$, and dust temperatures of $\sim 40$ K. 
All the derived physical parameters of the Norma cores, except the last, 
are similar to those of cores harboring UC HII regions (Hunter et al. 2000;
Fa\'undez et al. 2004). For a sample of about 150 IRAS point sources with 
colours of UC HII regions and average luminosity of $2.3\times10^5$ \Lsun,
Fa\'undez et al. (2004) derived an average mass of $5\times10^3$ M$_\odot$, 
an average size of 0.4 pc, and an average density of $2\times10^5$ cm$^{-3}$.
The dust temperature of the Norma cores are however larger than of the 
Fa\'undez et al. cores, for which they find an average temperature of 32 K. The 
larger values derived for the Norma cores are consistent with them being 
centrally heated by more luminous objects (average luminosity of 
$9.7\times10^5$ \Lsun).

We find that the observed radial intensity profiles of the dust continuum 
emission from the Norma cores are well fitted with single power-law profiles, 
with intensity indices in the range 1.5 to 1.9. This in turn indicates that the 
cores are highly centrally condensed, having radial density profiles with 
power-law density indices in the range $1.9-2.3$. These steep profiles are 
thought to indicate objects in the collapse and accretion phase of evolution.

We find that under the physical conditions of the Norma cores, high-mass stars 
can migrate towards the central region due to gravitational drag in time 
scales of $5\times10^{5}$~yr, smaller than the estimated ages of the clusters, 
providing an explanation for the observed stellar mass segregation within 
the cores (Paper II).

\acknowledgments

The authors gratefully acknowledge support from 
CONICYT through projects FONDAP No. 15010003 and BASAL PFB-06. 

\newpage

\begin{deluxetable}{lccccc}
\tablewidth{0pt}
\tablecaption{OBSERVED SOURCES\label{tbl-obssources}}
\tablehead{
\colhead{Source}          & \colhead{IRAS}            & 
\colhead{$\alpha$(2000)}  & \colhead{$\delta$(2000)}  & 
\colhead{D}          & \colhead{$\cal{L}_{\small IRAS}$}  \\
\colhead{ }          & \colhead{}              &
\colhead{ }          & \colhead{}              &
\colhead{(kpc)}      & \colhead{(\lo)}
}
\startdata
G324.201+0.119  & 15290-5546 &  $15^{\rm h}32^{\rm m}53\fs2$ & 
                          $-55\arcdeg56\arcmin13\arcsec$   & 6.0 & $5.9\times10^5$ \\
G328.307+0.423 & 15502-5302 & 15~~54~~06.3  & $-$53 11 38 & 5.6 & $1.1\times10^6$ \\
G329.337+0.147 & 15567-5236 & 16~~00~~33.4  & $-$52 44 45 & 7.0 & $1.2\times10^6$ \\
G330.949-0.174 & 16060-5146 & 16~~09~~49.1  & $-$51 54 45 & 5.4 & $1.0\times10^6$ \\
\enddata
\end{deluxetable}

\begin{deluxetable}{lcrccc}
\tablewidth{0pt}
\tablecaption{MOLECULAR LINES: OBSERVATIONAL PARAMETERS \label{tbl-obspar}}
\tablehead{
\colhead{Molecule}      & \colhead{Transition}          &
\colhead{Frequency}     & \colhead{Beam Size}           &
\colhead{$\eta$}        & \colhead{$\Delta v$}          \\ 
\colhead{}              & \colhead{}                    &
\colhead{(MHz)}         & \colhead{(FWHM \arcsec)}      &
\colhead{}              & \colhead{(\kms)}
}
\startdata
CS         & \Jjj{2}{1}\           &  97980.968 & 52 & 0.73 & 0.132 \\
           & \Jjj{3}{2}\           & 146969.049 & 34 & 0.66 & 0.085 \\
           & \Jjj{5}{4}\           & 244935.606 & 22 & 0.45 & 0.053 \\
C$^{34}$S  & \Jjj{3}{2}\           & 144617.147 & 34 & 0.66 & 0.087 \\
SiO        & \Jjj{2}{1}\           &  86846.998 & 57 & 0.75 & 0.144 \\
           & \Jjj{3}{2}\           & 130268.702 & 40 & 0.68 & 0.098 \\
CH$_3$OH   & \Jjkk{2}{1}{0}\ A$^+$  &  96741.42~ & 52 & 0.73 & 0.129 \\
           & \Jjkk{3}{2}{-1}\ E     & 145097.470 & 34 & 0.66 & 0.086 \\
CO         & \Jjj{1}{0}\           & 115271.204 & 45 & 0.70 & 0.111 \\
$^{13}$CO  & \Jjj{1}{0}\           & 110201.353 & 47 & 0.71 & 0.116 \\
C$^{18}$O  & \Jjj{1}{0}\           & 109782.160 & 47 & 0.71 & 0.116 \\

\enddata
\end{deluxetable}

\begin{deluxetable}{lccccccccccc}
\tablewidth{0pt}
\tablecaption{INTEGRATION TIMES AND SENSITIVITIES \label{tbl-tinrms}}
\tablehead{
\colhead{Line}   &  \multicolumn{2}{c}{G324.201} & &  
\multicolumn{2}{c}{G328.307} & & \multicolumn{2}{c}{G329.337}  & &
\multicolumn{2}{c}{G330.949} \\
\cline{2-3} \cline{5-6}
\cline{8-9} \cline{11-12}
\colhead{ }          &
\colhead{t$_{in}$} & \colhead{$\sigma$\tablenotemark{a}} & &
\colhead{t$_{in}$} & \colhead{$\sigma$} & & 
\colhead{t$_{in}$} & \colhead{$\sigma$} & &
\colhead{t$_{in}$} & \colhead{$\sigma$} \\
\colhead{} &
\colhead{(m)} & \colhead{(K)} & & 
\colhead{(m)} & \colhead{(K)} & &  
\colhead{(m)} & \colhead{(K)} & &
\colhead{(m)} & \colhead{(K)}
}
\startdata
CS\jj{2}{1}        &  3  & 0.12  & & 3  & 0.14  & & 6  & 0.064 & & 3  & 0.094 \\
CS\jj{3}{2}        &  4  & 0.065 & & 4  & 0.071 & & 4  & 0.065 & & 4  & 0.057 \\  
CS\jj{5}{4}        &  5  & 0.47  & & 3  & 0.11  & & 6  & 0.41  & & 5  & 0.44  \\  
C$^{34}$S\jj{3}{2} &  4  & 0.074 & & 4  & 0.068 & & 4  & 0.068 & & 4  & 0.049 \\ 
SiO\jj{2}{1}       & 20  & 0.018 & & 12 & 0.028 & & 12 & 0.027 & & 8  & 0.026 \\
SiO\jj{3}{2}       & 12  & 0.033 & & 22 & 0.016 & & 22 & 0.021 & & 4  & 0.035 \\
CH$_3$OH\jk{2}{1}  & 8   & 0.037 & & -- & --    & & 4  & 0.054 & & 4  & 0.038 \\
CH$_3$OH\jk{3}{2}  & 8   & 0.043 & & 8  & 0.045 & & 8  & 0.044 & & 4  & 0.049 \\
CO\jj{1}{0}        & 1   & 0.276 & & 1  & 0.276 & & 1  & 0.254 & & 1  & 0.254 \\
$^{13}$CO\jj{1}{0} & 3   & 0.072 & & 3  & 0.071 & & 3  & 0.068 & & 3  & 0.076 \\ 
C$^{18}$O\jj{1}{0} & 6   & 0.057 & & 6  & 0.055 & & 5  & 0.060 & & 5  & 0.063 \\ 
\enddata
\tablenotetext{a}{ 1$\sigma$ rms noise in antenna temperature.}
\end{deluxetable}

\begin{deluxetable}{lccccccc}
\tabletypesize{\small}
\tablewidth{0pt}
\small
\tablecolumns{8}
\tablecaption{MOLECULAR LINES: OBSERVED PARAMETERS \label{tbl-linepar}}
\tablehead{
\colhead{Line}   &  \multicolumn{3}{c}{Peak \tablenotemark{a}} & &  
\multicolumn{3}{c}{Average \tablenotemark{b}} \\ 
\cline{2-4} \cline{6-8}
\colhead{ }          &
\colhead{T$_A^*$} & \colhead{V} & \colhead{$\Delta$v} & & 
\colhead{T$_A^*$} & \colhead{V} & \colhead{$\Delta$v}  \\
\colhead{ }          &
\colhead{(K)} &  \colhead{(\kms)} &  \colhead{(\kms)} & &  
\colhead{(K)} &  \colhead{(\kms)} &  \colhead{(\kms)} 
}
\startdata
\cutinhead{G324.201+0.119}
CS\jj{2}{1}   & $1.79\pm$0.12 &  $-88.44\pm$.04 & $4.1\pm$0.3 
& & $0.537\pm$.02  & $-88.16\pm.03$ & $4.6\pm0.1$ \\ 
CS\jj{3}{2}   & $2.17\pm$0.07 &  $-88.52\pm$.01 & $4.1\pm$0.1  
& & $1.31\pm$.02   & $-88.34\pm.09$ & $4.4\pm0.1$ \\ 
CS\jj{5}{4}   & $2.98\pm$0.47 &  $-87.87\pm$.05 & $4.4\pm$0.1 
& & $2.33\pm$0.16  & $-88.33\pm.04$ & $3.8\pm0.1$ \\
C$^{34}$S\jj{3}{2} & $0.745\pm$0.074 &  $-88.59\pm$.05 & $3.6\pm$0.1
& & $0.276\pm.02 $ & $-88.33\pm.05$ & $3.4\pm0.2$ \\
SiO\jj{2}{1}  & $0.052\pm$0.018 &  $-87.4\pm$0.3 & $6.1\pm$1.1 
& & $0.025\pm.006 $ & $-88.74\pm.18$  & $3.4\pm0.5$ \\
SiO\jj{3}{2}  & --   & --   &  -- 
& & $0.017\pm.011 $ & $-88.6\pm0.3$  & $2.6\pm0.7$ \\ 
CH$_3$OH\jkk{2}{1}{0}A$^+$ & $0.380\pm$0.037 &  $-88.26\pm$.05 & $4.6\pm$0.2 
& & $0.270\pm.012 $ & $-88.19\pm.02$ & $4.2\pm0.1$ \\ 
CH$_3$OH\jkk{3}{2}{0}A$^+$ & $0.510\pm$0.043 &  $-88.36\pm$.03 & $5.5\pm$0.1 
& & $0.277\pm.014 $ & $-88.24\pm.02$ & $4.4\pm0.1$ \\ 
\cutinhead{G328.307+0.423} 
CS\jj{2}{1}   & -- & --$^c$ & --  
& & $0.858\pm0.028$ & $-92.65\pm.03$ & $6.3\pm0.1$  \\ 
CS\jj{3}{2}   & $2.93\pm0.07$ &  $-92.90\pm$.02 & $4.8\pm$0.1 
& & $1.84\pm$0.02 & $-92.92\pm.01$ & $4.3\pm0.1$  \\ 
CS\jj{5}{4}   & $1.18\pm$0.11 &  $-92.82\pm$.04 & $3.6\pm$0.2 
& & --  & --  & --  \\
C$^{34}$S\jj{3}{2}  & $0.45\pm0.068$ &  $-92.49\pm$.09 & $4.4\pm$0.2 
& & $0.386\pm$0.02 & $-92.32\pm.03$ & $5.3\pm0.1$  \\
SiO\jj{2}{1}  & $0.086\pm$0.028  &  $-92.24\pm$.13 & $6.5\pm$0.4  
& & --          &  --$^d$            & --           \\
SiO\jj{3}{2}  & $0.099\pm0.016$  &  $-92.75\pm$.10 & $6.1\pm$0.3 
& & --          &  --            & --           \\ 
CH$_3$OH\jkk{3}{2}{0}A$^+$ & $0.37\pm$0.045 &  $-91.84\pm$.05 & $5.4\pm$0.1 
& & $0.273\pm0.011 $ & $-92.71\pm.02$ & $4.9\pm0.1$  \\ 
\cutinhead{G329.337+0.147} 
CS\jj{2}{1}   & $3.34\pm$0.06 &  $-107.69\pm$.02 & $4.3\pm$0.1 
& & $1.70\pm$0.01 & $-107.58\pm.01$ & $5.2\pm0.1$ \\
CS\jj{3}{2}   & $4.92\pm$0.07 &  $-107.71\pm$.01 & $4.7\pm$0.1 
& & $1.81\pm$0.02 & $-107.57\pm.01$ & $4.5\pm0.1$ \\ 
CS\jj{5}{4}   & $2.23\pm$0.41 &  $-107.73\pm$.08 & $4.5\pm$0.3 
& & $1.91\pm$0.14 & $-107.76\pm.04$ & $4.6\pm0.2$  \\
C$^{34}$S\jj{3}{2}  & $1.15\pm$0.07 &  $-107.72\pm$.03 & $3.6\pm$0.1 
& & $0.246\pm$0.02 & $-107.66\pm.02$ & $3.0\pm0.1$  \\
SiO\jj{2}{1}  & $0.104\pm$0.027 &  $-107.67\pm$.10 & $4.2\pm$0.2 
& & $0.067\pm$0.007 & $-107.58\pm.05$ & $4.5\pm0.2$  \\
SiO\jj{3}{2}  & $0.125\pm0.021$ &  $-107.47\pm$.13 & $4.2\pm$0.3 
& & --          &  --            & --            \\  
CH$_3$OH\jkk{2}{1}{0}A$^+$ & $0.502\pm$0.054 & $-107.51\pm$.04 & $3.8\pm$0.1 
& & $0.355\pm$0.014 & $-107.50\pm.02$ & $4.0\pm0.1$  \\ 
CH$_3$OH\jkk{3}{2}{0}A$^+$ & $0.671\pm$0.044 &  $-107.48\pm$.02 & $4.8\pm$0.1 
& & $0.322\pm0.012 $ & $-107.48\pm.02$ & $4.5\pm0.1$ \\ 
\cutinhead{G330.949-0.174} 
CS\jj{2}{1}   & -- & --$^c$ & --      
& & $0.585\pm0.019$ & $-91.16\pm.02$ & $7.2\pm0.1$  \\ 
CS\jj{3}{2}   & -- & --$^c$ & --     
& & $1.80\pm0.016$ & $-91.65\pm.02$ & $7.2\pm0.1$  \\ 
CS\jj{5}{4}   & -- & --$^c$ & --     
& & $3.05\pm0.15$ & $-92.12\pm.03$ & $9.0\pm0.1$  \\
C$^{34}$S\jj{3}{2} & $0.624\pm$0.049 &  $-91.23\pm$.06 & $4.8\pm$0.5  
& & $0.270\pm0.014$ & $-91.01\pm.02$ & $4.3\pm0.2$   \\
SiO\jj{2}{1}  & $0.501\pm$0.026 &  $-90.24\pm$.03 & $6.2\pm$0.2 
& & $0.250\pm0.08$ & $-90.3\pm0.1$ & $6.0\pm0.1$    \\ 
SiO\jj{3}{2}  & $0.461\pm0.035$ &  $-90.56\pm$.01 & $6.1\pm$0.2 
& & --          &  --            & --            \\  
CH$_3$OH\jkk{2}{1}{0}A$^+$ & $0.985\pm0.038$ &  $-90.35\pm$.03 & $6.1\pm$0.1 
& & $0.538\pm0.011$ & $-90.3\pm0.1$ & $5.7\pm0.1$    \\ 
CH$_3$OH\jkk{3}{2}{0}A$^+$ & $1.24\pm0.049$ &  $-90.95\pm$.02 & $7.9\pm$0.1 
& & $0.549\pm0.014$ & $-90.7\pm0.1$ & $6.9\pm0.1$    \\ 
\enddata
\tablenotetext{a}{Fits to the spectra at peak position.}
\tablenotetext{b}{Fits to the spatially averaged emission.}
\tablenotetext{c}{Non-Gaussian profile.}
\tablenotetext{d}{Only broad emission detected in this line.}
\end{deluxetable}

\begin{deluxetable}{llccc}
\tablewidth{0pt}
\small
\tablecolumns{5}
\tablecaption{WING EMISSION: VELOCITY EXTENT \label{tbl-wingpar}}
\tablehead{
\colhead{Source}          & \colhead{Line}          &
\colhead{V$_{min}$} & \colhead{V$_{max}$} & \colhead{$\Delta$v} \\
\colhead{ }          &
\colhead{(\kms)} &  \colhead{(\kms)} &  \colhead{(\kms)}
}
\startdata
G324.201+0.119... & CS\jj{3}{2}    & $-$101.5 & $-$77.0 & 24.5 \\
                  & SiO\jj{2}{1}   & $-$101.0 & $-$71.6 & 29.4 \\
                  & SiO\jj{3}{2}      & $-$101.4 & $-$72.0 & 29.4 \\
G328.307+0.423... & CS\jj{3}{2}    & $-$101.0 & $-$80.6 & 20.4 \\
                  & SiO\jj{2}{1}   & $-$100.0 & $-$82.6 & 17.4 \\
G329.337+0.147... & CS\jj{3}{2}    & $-$122.5 & $-$90.9 &31.6 \\
                  & SiO\jj{2}{1}   & $-$114.0 & $-$92.2 &21.8 \\
                  & SiO\jj{3}{2}   & $-126.9$ & $-$90.3  & 36.6 \\
G330.949-0.174    & CS\jj{3}{2}    & $-$106.0 & $-$75.7 & 30.3 \\
                  & SiO\jj{2}{1}   & $-$107.0 & $-$70.4 & 36.6 \\
                  & SiO\jj{3}{2}   & $-$103.4 & $-$75.2 & 28.2 \\
\enddata
\end{deluxetable}

\begin{deluxetable}{lcccccc}
\tablewidth{0pt}
\tablecolumns{7}
\tablecaption{1.2-MM CONTINUUM EMISSION: OBSERVED PARAMETERS
  \label{tbl-obspardust}}
\tablehead{
\colhead{SIMBA}  & \multicolumn{2}{c}{Peak position} & \colhead{} &
 \multicolumn{2}{c}{Flux density\tablenotemark{a}} & \colhead{Angular size
\tablenotemark{b}} \\
\cline{2-3} \cline{5-6}
\colhead{source}                  & \colhead{$\alpha$(2000)}      &
\colhead{$\delta$(2000)}  &  & \colhead{Peak}  & \colhead{Total} &
\colhead{(\arcsec)}  \\
\colhead{}                  & \colhead{}      &
\colhead{} &   & \colhead{(Jy/beam)}  & \colhead{(Jy)} &
\colhead{} 
}
\startdata
G324.201+0.119  & 15 32 52.91 & -55 56 10.8  & & 5.28  & 17.3 &  $35\times28$ \\
G328.307+0.423  & 15 54 06.33 & -53 11 43.1  & & 6.35  & 24.1 &  $43\times26$ \\
G329.337+0.147A & 16 00 33.10 & -52 44 46.6  & & 4.35  & 11.7 &  $37\times27$ \\
G329.337+0.147B & 16 00 38.68 & -52 45 31.7  & & 1.45  &  4.58 & $53\times38$ \\
G330.949-0.174  & 16 09 52.41 & -51 54 58.5  & & 21.75 & 47.2  & $34\times33$ \\
\enddata
\tablenotetext{a}{Errors in the flux density are dominated by the
uncertainties in the flux calibration, of $\sim$20\%.}
\tablenotetext{b}{Errors in the angular sizes are typically 10\%.}
\end{deluxetable}

\begin{deluxetable}{lcccccc}
\tablewidth{0pt}
\tablecolumns{7}
\tablecaption{DUST EMISSION: DERIVED PARAMETERS \label{tbl-deriveddust}}
\tablehead{
\colhead{SIMBA} & \colhead{T$_d$} & \colhead{Radius} &
 \colhead{Mass} &  \colhead{n(H$_2$)}  & \colhead{N(H$_2$)} &
 \colhead{$\tau_{1.2mm}$} \\
 \colhead{source} & \colhead{(K)} & \colhead{(pc)} & \colhead{(M$_{\odot}$)} &
 \colhead{(cm$^{-3})$} & \colhead{(cm$^{-2}$)}  & \colhead{}
}
\startdata
G324.201+0.119 &  37. & 0.45 & $5.0\times10^3$ & $2.2\times10^5$
& $4.1\times10^{23}$ & 0.016 \\
G328.307+0.423 &  42. & 0.46 & $5.2\times10^3$ & $2.2\times10^5$
& $4.2\times10^{23}$ & 0.017 \\
G329.337+0.147A &  40. & 0.53 & $4.1\times10^3$ & $1.1\times10^5$
& $2.5\times10^{23}$ & 0.010 \\
G329.337+0.147B &  40. & 0.76 & $1.6\times10^3$ & $1.5\times10^4$
& $4.8\times10^{22}$ & 0.002 \\
G330.949-0.174 & 37. & 0.44 & $1.1\times10^4$ & $5.3\times10^5$
& $9.6\times10^{23}$ & 0.038  \\
\enddata
\end{deluxetable}

\begin{deluxetable}{lcccc}
\tablewidth{0pt}
\tablecolumns{5}
\tablecaption{MOLECULAR LINES: OPACITIES \label{tbl-opacities}}
\tablehead{
\colhead{Source}   & \multicolumn{4}{c}{Peak optical depth} \\
\cline{2-5}
\colhead{ }          & 
\colhead{CS(3$\rightarrow$2)} & \colhead{C$^{34}$S(3$\rightarrow$2)} & 
\colhead{CO(1$\rightarrow$0)} & \colhead{$^{13}$CO(1$\rightarrow$0)}
}
\startdata
G324.201+0.119 & 7.0 & 0.30 & 42 & 0.7       \\
G328.307+0.423 & 3.4 & 0.15 & $>26$ & $>0.4$ \\
G329.337+0.147 & 5.5 & 0.24 & 31 & 0.5       \\ 
G330.949-0.174 & 7.1 & 0.31 & 69 & 1.1       \\
\enddata
\end{deluxetable}

\begin{deluxetable}{lccccccccc}
\tabletypesize{\small}
\tablewidth{0pt}
\tablecolumns{10}
\tablecaption{COLUMN DENSITIES AND ROTATIONAL TEMPERATURES$^a$ \label{tbl-colden}}
\tablehead{
\colhead{Source}   & \multicolumn{6}{c}{Ambient gas} & &  
  \multicolumn{2}{c}{Outflow} \\
\cline{2-7}
\cline{9-10}
\colhead{ }  & \colhead{N(CS)}  & \colhead{N(CO)} & \colhead{N(CH$_3$OH)} 
   & \colhead{N(SiO)} & \colhead{T(CH$_3$OH)} & \colhead{T(SiO)} & \colhead{ } 
   & \colhead{N(SiO)} & \colhead{T(SiO)}  
}
\startdata
G324.201+0.119 & $7.6\times10^{15}$
& $1.7\times10^{19}$ & $9.7\times10^{14}$ & --- & 12.0 & --- & & 
  $1.3\times10^{13}$ & 31.0 \\
G328.307+0.423 & $5.2\times10^{15}$
& $1.4\times10^{19}$ & $9.4\times10^{14}$ & $4.7\times10^{12}$ & 10.3 & 6.3 & & 
  $3.8\times10^{12}$ & 24.3 \\
G329.337+0.147 & $5.8\times10^{15}$
& $1.4\times10^{19}$ & $9.0\times10^{14}$ & $6.3\times10^{12}$ & 17.0 & 16.7 & &
  $1.8\times10^{12}$ & 45.6 \\
G330.949-0.174 & $1.6\times10^{16}$
& $3.9\times10^{19}$ & $2.1\times10^{15}$ & $4.1\times10^{13}$ & 22.6 & 11.2 & & 
  $2.6\times10^{13}$ & 24.9 \\
\enddata
\tablenotetext{a}{N: column density in cm$^{-2}$; T: rotational temperature in K}
\end{deluxetable}

\begin{deluxetable}{lccc}
\tablewidth{0pt}
\tablecolumns{4}
\tablecaption{MOLECULAR LINES: DERIVED PARAMETERS OF CORES \label{tbl-derivedpar}}
\tablehead{
\colhead{Source}   &  \colhead{Radius}  & 
\multicolumn{2}{c}{Mass$^a$} \\ 
\cline{3-4} 
\colhead{ }          & \colhead{}   &
\colhead{M$_{\small CS}$}  & \colhead{M$_{vir}$} \\ 
\colhead{ }          & \colhead{(pc)}   &
\colhead{(\mo)}       & \colhead{(\mo)} 
}
\startdata
G324.201+0.119 & 0.52 & $4.3\times10^3$ &  $2.3\times10^3$  \\ 
G328.307+0.423 & 0.66 & $4.6\times10^3$ &  $5.5\times10^3$  \\ 
G329.337+0.147 & 0.71 & $6.0\times10^3$ &  $4.0\times10^3$  \\ 
G330.949-0.174 & 0.56 & $1.0\times10^4$ &  $6.1\times10^3$  \\ 
\enddata
\tablenotetext{a}{Derived from CS(2$\rightarrow$1) observations.}
\end{deluxetable}

\begin{deluxetable}{lcccc}
\tablewidth{0pt}
\tablecolumns{4}
\tablecaption{ABUNDANCE RATIOS \label{tbl-abundances}}
\tablehead{
\colhead{Source}  & \colhead{CS/CO} & \colhead{SiO/CO} & \colhead{CH$_3$OH/CO} \\
}
\startdata
G324.201+0.119  & $4.5\times10^{-4}$ &     ---            & $5.7\times10^{-5}$ \\
G328.307+0.423  & $3.7\times10^{-4}$ & $3.3\times10^{-7}$ & $6.7\times10^{-5}$ \\ 
G329.337+0.147  & $4.1\times10^{-4}$ & $4.5\times10^{-7}$ & $6.4\times10^{-5}$ \\ 
G330.949-0.174  & $4.1\times10^{-4}$ & $1.0\times10^{-6}$ & $5.4\times10^{-5}$ \\ 
\enddata
\end{deluxetable}


\clearpage


\clearpage

\clearpage

\clearpage

\clearpage

\end{document}